\ifx\onecol\undefined
\documentclass[journal,twocolumn]{IEEEtran}
\else 
\documentclass[12pt,draftclsnofoot,journal,onecolumn]{IEEEtran}
\fi

\IEEEoverridecommandlockouts

\usepackage{geometry}
\usepackage[noend]{algpseudocode}
\usepackage{balance}
\usepackage{algorithmicx,algorithm}
\usepackage{eqparbox}
\usepackage{amsmath}
\usepackage{subfigure}
\usepackage{pdfcolfoot}
\usepackage{verbatim}
\usepackage{braket}
\usepackage{cases}
\usepackage{amsthm,amsmath,amssymb,bm,bbm}

\newtheorem{lemma}{Lemma}

\newtheorem{remark}{Remark}
\newtheorem{corollary}{Corollary}
\usepackage{mathrsfs}

\usepackage{hyperref}
\hypersetup{hidelinks, 
colorlinks=true,
allcolors=black,
pdfstartview=Fit,
breaklinks=true}

\usepackage{graphicx}
\usepackage{epstopdf}
\usepackage{cite}
\makeatletter

\newcommand{\biggg}{\bBigg@{3}}

\newcommand{\Biggg}{\bBigg@{3.5}}

\newcommand{\bigggg}{\bBigg@{4}}

\newcommand{\Bigggg}{\bBigg@{4.5}}

\makeatother

\begin{document}
%
%
%
%


\title{Near-Field Channel Modeling for Electromagnetic Information Theory
}
\author{
	\IEEEauthorblockN{
		Zhongzhichao~Wan,
		Jieao~Zhu,
		and~Linglong~Dai,~\IEEEmembership{Fellow,~IEEE}
	}
	\thanks{This work was supported in part by the National Natural Science Foundation of China (Grant No. 62325106), in part by the National Key Research and Development Program of China (Grant No. 2023YFB3811503). }
	\thanks{All authors are with the Department of Electronic Engineering, Tsinghua University as well as Beijing National Research Center for Information Science and Technology (BNRist), Beijing 100084, China (E-mails: \{wzzc20, zja21\}@mails.tsinghua.edu.cn; daill@tsinghua.edu.cn).}

}

\maketitle



%

\begin{abstract}
	Electromagnetic information theory (EIT) is one of the emerging topics for 6G communication due to its potential to reveal the performance limit of wireless communication systems. For EIT, the research foundation is reasonable and accurate channel modeling. Existing channel modeling works for EIT in non-line-of-sight (NLoS) scenario focus on far-field modeling, which can not accurately capture the characteristics of the channel in near-field. In this paper, we propose the near-field channel model for EIT based on electromagnetic scattering theory. We model the channel by using non-stationary Gaussian random fields and derive the analytical expression of the correlation function of the fields. {Furthermore, we analyze the characteristics of the proposed channel model, e.g., channel degrees of freedom (DoF). Finally, we design a channel estimation scheme for near-field scenario by integrating the electromagnetic prior information of the proposed model.} Numerical analysis verifies the correctness of the proposed scheme and shows that it can outperform existing schemes like least square (LS) and orthogonal matching pursuit (OMP).
\end{abstract}


\section{Introduction}
To improve the system performance, various promising technologies, including reconfigurable intelligent surfaces (RISs) \cite{basar2019wireless,wang2022location}, continuous-aperture multiple-input multiple-output (CAP-MIMO) \cite{huang2020holographic,zhang2023pattern}, and near-field communications \cite{cui2022near,wu2023multiple}, have been recently investigated for sixth-generation (6G) communication. All these technologies try to explore new sources of degrees of freedom (DoF) or capacity gain for performance improvement. The performance gain actually come from more accurate understanding and precise manipulation of electromagnetic fields which convey information \cite{chafii2023twelve}. Therefore, combining classical electromagnetic theory and information theory to provide modeling and capacity analysis tools is of great importance for exploring the fundamental performance limit of wireless communication systems, which leads to the interdisciplinary subject called electromagnetic information theory (EIT) \cite{migliore2018horse}. By integrating deterministic physical theory and stochastic mathematically theory \cite{zhu2022electromagnetic}, EIT is expected to provide new insights into system models, degrees of freedom, capacity limits, etc., from the electromagnetic perspective. 

\subsection{Prior works}

The existing research directions of EIT includes channel modeling \cite{gong2023holographic,wei2023tri,pizzo2023wide}, DoF analysis \cite{bucci1987spatial, bucci1989degrees,franceschetti2015landau}, mutual information and capacity analysis \cite{jensen2008capacity,jeon2017capacity,wan2022mutual}, etc.
Among these directions, channel modeling is the fundamental part. Without precise channel model, DoF and capacity of EIT can not be accurately analyzed.

For the channel modeling schemes of EIT, one approach is line-of-sight (LoS) modeling scheme derived directly from Maxwell's equations, and the channel is expressed by the Green's function in free space \cite{gong2023holographic,wei2023tri} or considering reflection from a surface \cite{pizzo2023wide}. Another approach considers non-line-of-sight (NLoS) channel, which obeys the electromagnetic scattering theory \cite{chew1999waves}. Compared to LoS channel, NLoS channel is more general and can support larger degrees of freedom in wireless communication. Due to the complexity and uncertainty of the scattering environment, the NLoS channel is often modeled by random fields, whose statistical characteristics are derived from the scattering environment \cite{pizzo2022fourier}. For example, an isotropic statistical channel model was derived in \cite{bjornson2020rayleigh}, where correlation exists between different points of the received field. This model can be viewed as an extension of the traditional independent and
identically distributed (i.i.d.) Rayleigh fading channel model.
Furthermore, a more general scheme for constructing small-scale fading channel was provided in \cite{Marzetta'20}, where the received electromagnetic field was expanded by Fourier plane waves. {This model is based on the spatially stationary fields, where the correlation function of the random fields only depends on the distance vector between two points. Further work in \cite{pizzo2022spatial} discussed the MIMO model under the same assumption and simulated the capacity change with the antenna density \cite{pizzo2022fourier}.} An extended work \cite{demir2022channel} further derived approximate analytical correlation function based on \cite{Marzetta'20}, leading to a non-isotropic channel model. 

The above works have well analyzed the model of spatially stationary channel for EIT. The spatial stationarity is mathematically equivalent to the independence of the incoming waves at different angles. {In the far-field, each scatterer corresponds to a single direction for the incident waves on the receiver, which matches this assumption well. However, for the use of millimeter-wave and terahertz technologies and extremely large antenna aperture \cite{liu2023near}, such approximation is not accurate any more, where spatially non-stationary model can better capture the channel characteristics. Approximating the channel model in these scenarios by spatially stationary fields may introduce non-negligible errors in channel estimation, pattern design, capacity analysis, etc.} Therefore, an accurate channel modeling scheme for the EIT in NLOS scenarios is needed.

\subsection{Our contributions}
Different from the existing works, in this paper, we propose a near-field spatially non-stationary channel model and the corresponding channel estimation scheme for EIT\footnote{Simulation codes are provided to reproduce the results in this paper: \url{http://oa.ee.tsinghua.edu.cn/dailinglong/publications/publications.html}.}.  Specifically, the contributions of this paper are summarized as follows:

\begin{itemize}
	\item{We propose a near-field channel model for EIT based on the electromagnetic scattering theory. The channel is modeled by zero-mean Gaussian random fields. Then its correlation function can fully describe the channel. An approximate analytical expression of the correlation function of the channel is derived, and its approximation accuracy is verified by numerical simulation.}
	\item{We analyze the characteristics of the proposed near-field channel model. We show how to generate one sample of the random field channel model. {Then, we show the fitness of our analytical model to the channel generated by well-accepted clustered delay line (CDL) model by using quasi-Newton algorithm. 
		Finally, we analyze how the parameters of the model affect the accuracy and degree of freedom (DoF) of the channel.} }
	\item{{We design a channel estimation scheme by integrating the electromagnetic prior information of the proposed channel model for channel estimation.} Numerical simulations show that the designed channel estimation scheme outperforms existing schemes like LS and OMP. }	
\end{itemize}

\subsection{Organization and notation}

\emph{Organization}: The rest of this paper is organized as follows. In Section. \ref{Sec_EMmodel}, we provide the electromagnetic model of NLoS channel based on the electromagnetic scattering theory. Then, in Section. \ref{Sec_analytical_model}, we derive an approximated analytical expression of the correlation function of the channel based on the electromagnetic model. In Section. \ref{sec_charac}, we analyze the characteristics of the proposed model, including fitness to CDL model, DoF, etc. Based on the proposed model, in Section. \ref{Sec_estimation} we design a near-field channel estimation scheme and verify its correctness by numerical simulations. Finally, in Section. \ref{Sec_conclusion} we provide the conclusions and possible future directions of our work.

\emph{Notation}: bold uppercase characters denote matrices;
bold lowercase characters denote vectors;
the dot $\cdot$ denotes the scalar product of two vectors, or the matrix-vector multiplication. 
${\mathbb E}\left[x\right]$ denotes the mean of random variable $x$; 
$\epsilon_0$ is the permittivity of a vacuum, $\mu_0$ is the permeability of a vacuum, and $c$ is the speed of light in a vacuum; 
$*$ denotes the convolution operation, and $\mathscr{F}[f(x)]$ denotes the Fourier transform of $f(x)$; 
$J_m(x)$ is the $m_{\rm th}$ order Bessel function of the first kind; $I_m(x)$ is the $m_{\rm th}$ order modified Bessel function. $\lfloor x \rfloor$ represents rounding $x$ down; $x \% y$ represents modulo operation.
$\det(\cdot)$ denotes the matrix determinant or the Fredholm determinant; {$\mathcal{A}_S$ represents the area of $S$}.

\section{Electromagnetic Model for Scattering Field}
\label{Sec_EMmodel}
Maxwell's equations are the fundamental physical laws of the electromagnetic system. For the characteristics of the scattering system, we can consider the scatterers as spatial non-uniformity of the electromagnetic characteristics like permittivity $\epsilon$ and permeability $\mu$. {We adopt the time-harmonic assumption which assumes that the electromagnetic waves oscillate on a single frequency point. Then we have ${\bf E}({\bf r},t) = {\bf E}({\bf r})e^{-{\rm j}\omega t}$, and the partial derivative $\partial/\partial t$ can be replaced by $-{\rm j}\omega$ \cite{gruber2008new}.
From the Maxwell's equations, we have}
\begin{subequations}
	\begin{align} 
		&\nabla \times {\bf E}({\bf r}) = {\rm j}\omega \mu({\bf r}) {\bf H}({\bf r}), \label{maxwell_1}\\
		& \nabla \times {\bf H}({\bf r}) = -{\rm j}\omega \epsilon({\bf r}) {\bf E}({\bf r})+{\bf J}({\bf r}),  \label{maxwell_2}\\
		& \nabla \cdot  (\epsilon({\bf r}) {\bf E}({\bf r})) = \rho({\bf r}),\\
		& \nabla \cdot (\mu({\bf r}) {\bf H}({\bf r})) = 0,
	\end{align}
\end{subequations}
where $\epsilon({\bf r})$, ${\mu}({\bf r})$ and $\rho({\bf r})$ represents the permittivity, permeability and charge density at the position ${\bf r}$. In homogeneous media, $\epsilon({\bf r})$ and ${\mu}({\bf r})$ will be constant, which is often used in light-of-sight channel modeling in the free space. Now we are considering inhomogeneous media, which can express the electromagnetic characteristics of scattering fields \cite{chew1999waves}. 

{
	By performing $\nabla\times$ on (\ref{maxwell_1}), we have
	\begin{equation}
		\nabla \times \mu({\bf r})^{-1} \nabla \times {\bf E}({\bf r}) = \nabla \times ({\rm j}\omega{\bf H}({\bf r})),
	\end{equation}
	 which leads to the corresponding vector wave equation
	 \begin{equation}
	 	\nabla \times \mu({\bf r})^{-1} \nabla \times {\bf E}({\bf r}) - \omega^2 \epsilon({\bf r})  {\bf E}({\bf r}) = {\rm j}\omega {\bf J}({\bf r}),
	 \end{equation}
	 where $k({\bf r}) = \omega \sqrt{\mu({\bf r})\epsilon({\bf r})}$ represents the inhomogeneous media over a finite domain $V$ according to \cite{chew1999waves}, and ${\rm j}\omega {\bf J}({\bf r})$ represents the source field. Outside the domain $V$, the wavenumber $k({\bf r})$ equals $k_0 = \omega\sqrt{\mu_0\epsilon_0}$.
	 By subtracting $\nabla \times \mu_0 \nabla  \times {\bf E}({\bf r})- \omega^2 \epsilon_0 {\bf E}({\bf r})$ from both sides and applying the Green's function, the received electric field can be expressed by
\begin{equation}
	\begin{aligned} 
		{\bf E}({\bf r}) &= {\rm j}\omega\int_{V_s} {\bf G}({\bf r},{\bf r}')\mu_0{\bf J}({\bf r}'){\rm d}{\bf r}' 
		\\&+ \int_{V} {\bf G}({\bf r},{\bf r}')(k^2({\bf r}')-k_0^2) 
		{\bf E}({\bf r}'){\rm d}{\bf r}',
	\end{aligned}
	\label{equ_scattering_field}
\end{equation}
where $V_s$ is the source region which generates the signal, ${\bf E}({\bf r}')$ is the induced electric field in the inhomogeneous regions in the space, and the dyadic Green's function ${\bf G}$ is the solution of the equation
\begin{equation}
	\begin{aligned} 
		\nabla \times \mu_0^{-1} \nabla \times {\bf G}({\bf r},{\bf r}') - \omega^2 \epsilon_0  {\bf G}({\bf r},{\bf r}') = \mu_0^{-1} {\bf I} \delta({\bf r}-{\bf r}').
	\end{aligned}
\end{equation}
Then, we can obtain the Green's function as ${\bf G}({\bf r},{\bf r}') = \frac{1}{{4\pi }} \left( {{\bf{I}} + \frac{{{\nabla _{\bf{r}}}\nabla _{\bf{r}}^{\rm{H}}}}{{{\kappa_0 ^2}}}} \right) \frac{{{e^{{\rm{j}}\kappa_0 \left\| {{\bf{r}} - {\bf{r}'}} \right\|}}}}{{\left\| {{\bf{r}} - {\bf{r}'}} \right\|}} $, which can be further expressed by \cite{poon2005degrees}:
\begin{equation}
	\begin{aligned}
		{\bf{G}}({\bf{r}},{\bf{r}'}) =& \frac{1}{{4\pi }}\frac{{{e^{{\rm{j}}\kappa_0 \left\| {{\bf{r}} - {\bf{r}'}} \right\|}}}}{{\left\| {{\bf{r}} - {\bf{r}'}} \right\|}}\Bigg[\left( {{\bf{I}} - {\bf{\hat p}}{{{\bf{\hat p}}}^{\rm{H}}}} \right) \\&+ \frac{{\rm j}}{2\pi \left\| {{\bf{r}} - {\bf{r}'}} \right\| /\lambda}\left( {\bf I}-3{\bf{\hat p}}{{{\bf{\hat p}}}^{\rm{H}}} \right) \\&-\frac{1}{(2\pi\left\| {{\bf{r}} - {\bf{r}'}} \right\|/\lambda )^2 }\left( {\bf I}-3{\bf{\hat p}}{{{\bf{\hat p}}}^{\rm{H}}}  \right) \Bigg] [{\rm m}^{-1}],
		\label{Green}
	\end{aligned}
\end{equation}
where ${\bf{\hat p}} = \frac{{\bf{r}} - {\bf{r}'}}{{\left\| {\bf{r}} - {\bf{r}'} \right\|}}$.
Here we assume that $\left\| {{\bf{r}} - {\bf{r}'}} \right\|/\lambda \gg 1$, which means that the receiver is in far-field of the scatterer's microstructure and holds true in general wireless communication scenarios \cite{danufane2021path}. Then, we can omit the items containing powers of $\frac{1}{\left\| {{\bf{r}} - {\bf{r}'}} \right\|/\lambda}$. Since ${\rm tr}({\bf I}-{\bf{\hat p}}{{{\bf{\hat p}}}^{\rm{H}}})({\bf I}-{\bf{\hat p}}{{{\bf{\hat p}}}^{\rm{H}}})^{\rm H} = 2$ is a constant, the average power of the electromagnetic field does not depend on the direction ${\bf{\hat p}}$ if the energy of the source current is equally distributed in all polarization directions. For simplicity, in this paper we reduce the vector wave field to scalar wave field showing the power of electric field averaged on all polarization directions. Physically if we consider the electromagnetic fields on a specific polarization direction $\hat{\bf a}$, an extra factor $\hat{\bf a}({\bf{I}} - {\bf{\hat p}}{{{\bf{\hat p}}}^{\rm{H}}})\hat{\bf a}^{\rm H}$ should be added. Under the scalar wave field, we have
\begin{equation}
	\begin{aligned} 
		E({\bf r}) =& {\rm j}\omega\mu_0\int_{V_s} g({\bf r},{\bf r}')J({\bf r}'){\rm d}{\bf r}'  
		\\& + \int_{V} g({\bf r},{\bf r}')(k^2({\bf r}')-k_0^2)E({\bf r}'){\rm d}{\bf r}',
	\end{aligned}
	\label{equ_scattering_field_scalar}
\end{equation}		
where $g({\bf r},{\bf r}') = \frac{1}{2\pi}\frac{e^{{\rm j}k_0\| {\bf r}-{\bf r}' \|}}{\|{\bf r}-{\bf r}' \|}$.
}
The equation (\ref{equ_scattering_field_scalar}) is an extension from the 2-dimensional case in \cite{li2018deepnis}. Here we can view the first item in (\ref{equ_scattering_field}) as the line-of-sight component of the field which is fixed and well-studied. Then we will focus on the second item in (\ref{equ_scattering_field}) which highly relies on the characteristics of the inhomogeneity of the space. The inhomogeneity of the space depends on the complicated factors such as surface structure and material properties of the medium which are hard to analytically model and may change over time. Therefore, a statistical model will be more suitable to depict the characteristics of the field than deterministic modeling scheme.

\section{Channel Model Based on Non-Stationary Random Fields}
\label{Sec_analytical_model}
\ifx\onecol\undefined
\begin{figure}
	\centering 
	\includegraphics[width=0.5\textwidth]{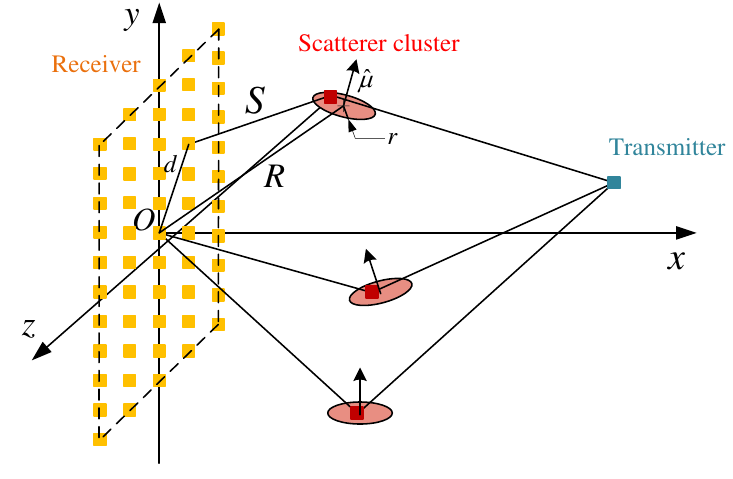} 
	\caption{The three-dimensional near-field statistical channel modeling where the scatterers are located in solid circles.} 
	\label{fig_3d_solid_sphere}
\end{figure}
\else
\begin{figure}
	\centering 
	\includegraphics[width=0.7\textwidth]{figs/threedimension_solid_circle.pdf} 
	\caption{The three-dimensional near-field statistical channel modeling where the scatterers are located in solid circles.} 
	\label{fig_3d_solid_sphere}
\end{figure}
\fi

In this section, we will derive the channel model for EIT based on the electromagnetic scattering theory explained in the above section. The channel is modeled as Gaussian random fields to tolerate the uncertainty of the inhomogeneity of the space \cite{franceschetti2006scattering}.
In our model the received field can be viewed as weighted superposition of spherical waves other than plane waves in \cite{Marzetta'20}. Therefore, it is suitable for both near-field and far-field communications by considering distances between antenna array and scatterers besides the azimuth and elevation angles.  

\subsection{Mathematical derivation of the analytical model}
By omitting the first item in (\ref{equ_scattering_field}) which represents the deterministic line-of-sight component, we have 
\begin{equation}
	\begin{aligned}
	R_E({\bf r}_1,{\bf r}_2) =& \mathbb{E} \Big[\int_V\int_V g({\bf r}_1,{\bf r}_1')g^{*}({\bf r}_2,{\bf r}_2')(k^2({\bf r}'_1)-k_0^2)
	\\& (k^2({\bf r}'_2)-k_0^2) E({\bf r}_1') E^{*}({\bf r}_2') {\rm d}{\bf r}_1'{\rm d}{\bf r}_2'\Big],
	\end{aligned}
	\label{equ_correlation_complicated}
\end{equation}
{where in the rest part of the paper we express $R({\bf r}_1,{\bf r}_2)$ as the abbreviation of $R_E({\bf r}_1,{\bf r}_2)$.}
To derive a closed-form expression of the channel model we need to have some assumptions on the scattering field to do simplifications on (\ref{equ_correlation_complicated}). { Considering the uncertainty of the inhomogeneity of the space, simplified models need to be used for the field correlation on the scatterer surfaces for ease of analysis. In the literature, some models have already been proposed. For example, delta function type of field spatial correlation was discussed in \cite{cabayan1973scattering}, \cite{osgood1999x} and \cite{dainty1977statistics}, which implies an ideal but mathematically friendly assumption that the material properties are varying instantaneously in the spatial domain. A more complex model like exponential type of field spatial correlation was discussed in \cite{tang1996regions}. Angular delta function as the spectrum of the field was proposed in \cite{ticconi2011models}. For simplicity, we adopt delta function as the spatial correlation of the fields on the scattering surface, leading to $\mathbb{E}\left[ E({\bf r}_1')E^{*}({\bf r}_2') \right] = \beta \delta({\bf r}_1'-{\bf r}_2')$.} Then, the channel correlation function reduces to 
\begin{equation}
	\begin{aligned}
	R({\bf r}_1,{\bf r}_2) =& \beta \int_V g({\bf r}_1,{\bf r}')g^{*}({\bf r}_2,{\bf r}')(k^2({\bf r}')-k_0^2)^2 {\rm d}{\bf r}'.
	\end{aligned}
	\label{equ_correlation_simplified}
\end{equation}

We further assume that the scattering region $V$ is distributed in a solid circle, centering at ${\bf d}$ and perpendicular to $\hat{\boldsymbol{\mu}}$, which means that $\hat{\boldsymbol{\mu}}^{\rm T}({\bf r}'-{\bf d})=0 $. {Here $\hat{\boldsymbol{\mu}}$ represents the direction of the scattering surface.} The practical meaning of such assumption is that the scatterer faces receivers at a certain angle. The radius of the circle is $r_s$. For the item $(k^2({\bf r}')-k_0^2)^2$, we view it as the gain of electromagnetic waves reflected from the surface of scatterer. Specifically, we model it by 
\begin{equation}
	f({\bf r}') = (k^2({\bf r}')-k_0^2)^2  =\left\{\begin{matrix}
	\frac{a+1}{\pi r_s^{2a+2}} (r_s^2-\rho^2)^{a}	&\left| \rho \right|\leqslant r_s, \\ 
	0	& {\rm otherwise},
	\end{matrix}\right.
\end{equation}
where $\boldsymbol{\rho} = {\bf r}' - {\bf d}$, $\rho = \|\boldsymbol{\rho}\|$, and $a$ is a parameter characterizing the concentration of scatterer around the central point. { This assumption for $f({\bf r}')$ is heuristic, which aims at providing a general model to cover different shapes of scatterers. By changing the parameter $a$, the scatterer varies from ring to single point.
For example, when $a = 0$, the scattering region is a uniform circular surface used in \cite{christou2016far}. When $a \rightarrow -1$, the scatterer approximates a ring as in \cite{mittal2009angle}. When $a \rightarrow +\infty$, the scattering region shrinks to a single point, which is widely adopted in existing works for near-field \cite{lu2023near}.} Then, we can express the correlation function by
\begin{equation}
	R({\bf r}_1,{\bf r}_2) = \beta \int_V \frac{e^{{\rm j}k\| {\bf r}_1-{\bf r}' \|}}{4\pi \|{\bf r}_1-{\bf r}' \|}\frac{e^{-{\rm j}k\| {\bf r}_2-{\bf r}' \|}}{4\pi \|{\bf r}_2-{\bf r}' \|} f({\bf r}'){\rm d}{\bf r}'.
\end{equation}

\begin{figure}
	\centering 
	\includegraphics[width=0.4\textwidth]{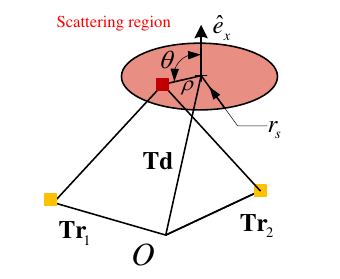} 
	\caption{The rotated coordinate system with ${\bf T}\hat{\boldsymbol{\mu}} = \hat{\bf e}_x$.} 
	\label{fig_rotation}
\end{figure}

To facilitate the derivation procedure, we choose a coordinate rotation ${\bf T}$ which satisfies ${\bf T}\hat{\boldsymbol{\mu}} = \hat{\bf e}_x$. Then we have a new rotated coordinate where $\boldsymbol{\mu}$ is the $x$ axis. The center of the scatterer is located at ${\bf Td}$, and the receiving locations are ${\bf T}{\bf r}_1$ and ${\bf T}{\bf r}_2$. One point in the scattering region is located at ${\bf Td}+{\bf T}\boldsymbol{\rho}$, where ${\bf T}\boldsymbol{\rho} = [\rho \cos\theta, \rho \sin\theta,0]$. Here we denote two directions  $\hat{\boldsymbol{\mu}}_1$ and $\hat{\boldsymbol{\mu}}_2$ perpendicular to $\hat{\boldsymbol{\mu}}$, which satisfies $\hat{\boldsymbol{\mu}}_1^{\rm T}\hat{\boldsymbol{\mu}}_2=0$. Then we can denote ${\bf Td}$ by $[{\bf d}\cdot \hat{\boldsymbol{\mu}},{\bf d}\cdot \hat{\boldsymbol{\mu}}_1,{\bf d}\cdot \hat{\boldsymbol{\mu}}_2]$. Similarly, we have ${\bf T}{\bf r} = [{\bf r}\cdot \hat{\boldsymbol{\mu}},{\bf r}\cdot \hat{\boldsymbol{\mu}}_1,{\bf r}\cdot \hat{\boldsymbol{\mu}}_2]$. The point in the scattering point is located at $[{\bf d}\cdot \hat{\boldsymbol{\mu}},{\bf d}\cdot \hat{\boldsymbol{\mu}}_1+\rho \cos \theta,{\bf d}\cdot \hat{\boldsymbol{\mu}}_2+ \rho \sin \theta]$. The rotated coordinate system is shown in Fig. \ref{fig_rotation}.
 The distance between ${\bf r}$ and ${\bf r}'$ is 
\ifx\onecol\undefined
(\ref{fig_S1})
\begin{figure*}
\begin{equation}
	\begin{aligned}
\| {\bf r}-{\bf r}'  \| &= \sqrt{\left( {\bf d}\cdot \hat{\boldsymbol{\mu}}- {\bf r}\cdot \hat{\boldsymbol{\mu}}\right)^2 + \left( {\bf d}\cdot \hat{\boldsymbol{\mu}}_1+\rho \cos \theta - {\bf r}\cdot \hat{\boldsymbol{\mu}}_1 \right)^2 + \left( {\bf d}\cdot \hat{\boldsymbol{\mu}}_2+ \rho \sin \theta - {\bf r}\cdot \hat{\boldsymbol{\mu}}_2 \right)^2}
\\&= d \sqrt{A({\bf r}) + 2\frac{\rho}{d}B({\bf r},\hat{\boldsymbol{\rho}})+  \left( \frac{\rho}{d} \right)^2},
\end{aligned}
\label{fig_S1}
\end{equation}
{\noindent} \rule[-10pt]{18cm}{0.05em}
\end{figure*}
\else
\begin{equation}
	\begin{aligned}
		S({\bf d}) &= \sqrt{\left( {\bf R}\cdot \hat{\boldsymbol{\mu}}- {\bf r}\cdot \hat{\boldsymbol{\mu}}\right)^2 + \left( {\bf R}\cdot \hat{\boldsymbol{\mu}}_1+\rho \cos \theta - {\bf r}\cdot \hat{\boldsymbol{\mu}}_1 \right)^2 + \left( {\bf R}\cdot \hat{\boldsymbol{\mu}}_2+ \rho \sin \theta - {\bf r}\cdot \hat{\boldsymbol{\mu}}_2 \right)^2}
		\\&= R \sqrt{A({\bf r}) + 2\frac{\rho}{R}B({\bf r},{\bf d})+  \left( \frac{\rho}{R} \right)^2},
\end{aligned}
\end{equation}
\fi
where
\ifx\onecol\undefined
\begin{equation}
	\begin{aligned}
A({\bf r}) =& 1+\left( \frac{r}{d} \right)^2 - 2\frac{r}{d}\Bigg( (\hat{\bf d}\cdot \hat{\boldsymbol{\mu}})(\hat{\bf r}\cdot \hat{\boldsymbol{\mu}}) \\&+ (\hat{\bf d}\cdot \hat{\boldsymbol{\mu}}_1)(\hat{\bf r}\cdot \hat{\boldsymbol{\mu}}_1)+(\hat{\bf d}\cdot \hat{\boldsymbol{\mu}}_2)(\hat{\bf r}\cdot \hat{\boldsymbol{\mu}}_2) \Bigg),
\end{aligned}
\end{equation}
\else
\begin{equation}
	\begin{aligned}
A({\bf r}) = 1+\left( \frac{r}{d} \right)^2 - 2\frac{r}{d}\left( (\hat{\bf d}\cdot \hat{\boldsymbol{\mu}})(\hat{\bf r}\cdot \hat{\boldsymbol{\mu}}) + (\hat{\bf d}\cdot \hat{\boldsymbol{\mu}}_1)(\hat{\bf r}\cdot \hat{\boldsymbol{\mu}}_1)+(\hat{\bf d}\cdot \hat{\boldsymbol{\mu}}_2)(\hat{\bf r}\cdot \hat{\boldsymbol{\mu}}_2) \right),
\end{aligned}
\end{equation}
\fi
and
\ifx\onecol\undefined
\begin{equation}
	\begin{aligned}
B({\bf r},\hat{\boldsymbol{\rho}}) =&  \hat{\bf d}\cdot \hat{\boldsymbol{\mu}}_1 \cos \theta + \hat{\bf d}\cdot \hat{\boldsymbol{\mu}}_2 \sin \theta - \frac{r}{d}\hat{\bf r} \cdot \hat{\boldsymbol{\mu}}_1 \cos \theta
 \\&-\frac{r}{d}\hat{\bf r} \cdot \hat{\boldsymbol{\mu}}_2 \sin \theta.
\end{aligned}
\end{equation}
\else
\begin{equation}
	\begin{aligned}
B({\bf r}) =  \hat{\bf d}\cdot \hat{\boldsymbol{\mu}}_1 \cos \theta + \hat{\bf d}\cdot \hat{\boldsymbol{\mu}}_2 \sin \theta - \frac{r}{d}\hat{\bf r} \cdot \hat{\boldsymbol{\mu}}_1 \cos \theta
 -\frac{r}{d}\hat{\bf r} \cdot \hat{\boldsymbol{\mu}}_2 \sin \theta.
\end{aligned}
\end{equation}
\fi

Through mathematical derivations and simplifications, we can derive the spatial correlation function of the channel in the following lemma, {where scatterer dimension is relatively small compared to the distance}:
\begin{lemma}[{Correlation function of the channel in weak near-field}]
	Assuming that $r_s \ll d$, the correlation function of the channel can be approximated by 
	\begin{equation}
		\begin{aligned}
			\tilde{R}({\bf r}_1,{\bf r}_2) =& \frac{\beta}{8\pi^2d^2\sqrt{A({\bf r}_1)A({\bf r}_2)}}e^{{\rm j}\frac{2\pi}{\lambda}R \left(\sqrt{A({\bf r}_1)}-\sqrt{A({\bf r}_2)}\right)} \\&(a+1) 2^{a} \Gamma (a+1) (\sqrt{C}r_s)^{-(a+1)} J_{a+1} (\sqrt{C}r_s),
		\end{aligned}
	\end{equation}
	where
	\ifx\onecol\undefined
   \begin{equation}
	   \begin{aligned}
		   C =& \left( \frac{2\pi}{\lambda}  \right)^2 \Bigg( \frac{\hat{\bf d}\cdot \hat{\boldsymbol{\mu}}_1}{\sqrt{A({\bf r}_1)}} - \frac{\hat{\bf d}\cdot \hat{\boldsymbol{\mu}}_1}{\sqrt{A({\bf r}_2)}} - \frac{r_1}{d} \frac{\hat{\bf r}_1 \cdot \hat{\boldsymbol{\mu}}_1}{\sqrt{A({\bf r}_1)}} \\&~~~~~~~~~~~~+ \frac{r_2}{R} \frac{\hat{\bf r}_2 \cdot \hat{\boldsymbol{\mu}}_1}{\sqrt{A({\bf r}_2)}}\Bigg)^2 \\&+  \left( \frac{2\pi}{\lambda}  \right)^2 \Bigg( \frac{\hat{\bf d}\cdot \hat{\boldsymbol{\mu}}_2}{\sqrt{A({\bf r}_1)}} - \frac{\hat{\bf d}\cdot \hat{\boldsymbol{\mu}}_2}{\sqrt{A({\bf r}_2)}} - \frac{r_1}{d} \frac{\hat{\bf r}_1 \cdot \hat{\boldsymbol{\mu}}_2}{\sqrt{A({\bf r}_1)}} \\&~~~~~~~~~~~~+ \frac{r_2}{d} \frac{\hat{\bf r}_2 \cdot \hat{\boldsymbol{\mu}}_2}{\sqrt{A({\bf r}_2)}}\Bigg)^2.
	   \end{aligned}
   \end{equation}
   \else
   \begin{equation}
	   \begin{aligned}
		C =& \left( \frac{2\pi}{\lambda}  \right)^2 \Bigg( \frac{\hat{\bf d}\cdot \hat{\boldsymbol{\mu}}_1}{\sqrt{A({\bf r}_1)}} - \frac{\hat{\bf d}\cdot \hat{\boldsymbol{\mu}}_1}{\sqrt{A({\bf r}_2)}} - \frac{r_1}{d} \frac{\hat{\bf r}_1 \cdot \hat{\boldsymbol{\mu}}_1}{\sqrt{A({\bf r}_1)}} \\&~~~~~~~~~~~~+ \frac{r_2}{R} \frac{\hat{\bf r}_2 \cdot \hat{\boldsymbol{\mu}}_1}{\sqrt{A({\bf r}_2)}}\Bigg)^2 \\&+  \left( \frac{2\pi}{\lambda}  \right)^2 \Bigg( \frac{\hat{\bf d}\cdot \hat{\boldsymbol{\mu}}_2}{\sqrt{A({\bf r}_1)}} - \frac{\hat{\bf d}\cdot \hat{\boldsymbol{\mu}}_2}{\sqrt{A({\bf r}_2)}} - \frac{r_1}{d} \frac{\hat{\bf r}_1 \cdot \hat{\boldsymbol{\mu}}_2}{\sqrt{A({\bf r}_1)}} \\&~~~~~~~~~~~~+ \frac{r_2}{d} \frac{\hat{\bf r}_2 \cdot \hat{\boldsymbol{\mu}}_2}{\sqrt{A({\bf r}_2)}}\Bigg)^2.
	   \end{aligned}
   \end{equation}
   \fi
	\end{lemma}
	\begin{IEEEproof}
	See Appendix A.
	\end{IEEEproof}

For the channel with multiple scatterers, the correlation function can be expressed by
\begin{equation}
	\begin{aligned}
		R({\bf r}_1,{\bf r}_2) = \sum_{k=1}^{M}\tilde{R}_k({\bf r}_1,{\bf r}_2),
	\end{aligned}
	\label{fig_R_multiple}
\end{equation}		
where each $\tilde{R}_k({\bf d}_1,{\bf d}_2)$ is constructed according to {\bf Lemma 1}. {For the channel with a large scatterer, we can decompose it to several small scatterers and express the channel in the form of (\ref{fig_R_multiple}).}

{
\begin{remark}
	In this paper before we arrive at {\bf Lemma 1}, several assumptions and simplifications are provided to reduce the analysis complexity and facilitate the derivation of an analytical result. These assumptions and simplifications include that 1) spatially uncorrelated scattered field adopted in \cite{cabayan1973scattering}, \cite{osgood1999x} and \cite{dainty1977statistics}, which implies an ideal but mathematically friendly assumption that the material properties are varying instantaneously in the spatial domain; 2) scattering region and gain function both for analysis convenience and generality, which covers different scatterer shapes like circle surface \cite{christou2016far}, ring \cite{mittal2009angle}, and single point \cite{lu2023near}; 3) distance is far larger than dimension of the scatterer. For larger scatterer, combinations of subchannels with small scatterers may provide an acceptable channel model; 4) distance is far larger than wavelength \cite{danufane2021path}; 5) scalar electromagnetic fields as in \cite{Marzetta'20}. By changing these assumptions or discarding these simplifications, a more accurate and general model may be obtained. 
	\end{remark}}

\subsection{Numerical verification of the accuracy of the analytical model}

In this subsection, we will show the accuracy of the analytical correlation function in {\bf Lemma 1}. We set the direction of the scattering region to the center of the array as $\hat{\bf d}=[\frac{1}{\sqrt{3}},\frac{1}{\sqrt{3}},\frac{1}{\sqrt{3}}]$. The scattering region is perpendicular to the direction $\hat{\boldsymbol{\mu}}=[-\frac{1}{\sqrt{3}},\frac{1}{\sqrt{3}},-\frac{1}{\sqrt{3}}]$. The concentration parameter on the scatterer cluster is set to $a=0$, which corresponds to uniform distribution on the circle. For the correlation between the received fields at two positions on the receiving array, we fix one position at the center of the array and another position at $[0,\pm n_yd_y,\pm n_zd_z]\,{\rm m}$, where $n_y,n_z \in \mathcal{I}_N$, $\mathcal{I}_N = \{ 1, \cdots, 100 \}$, $d_y=d_z = 0.025\,{\rm m}$. The wavelength $\lambda$ is set to $0.05\,{\rm m}$. We plot $\frac{\| \tilde{\bf R}-{\bf R}\|_{\rm F}^2}{\|{\bf R} \|_{\rm F}^2}$, which is the relative error between the approximated correlation matrix and the accurate correlation matrix, in Fig.~\ref{fig_approximation_error}. We can find that the approximation error is negligible compared to the value of the corresponding correlation function when $d$ is large enough or $r_s$ is small enough. For example, when $d$ is larger than $100\,{\rm m}$ and $r_s$ is smaller than $3.5\,{\rm m}$, the relative approximation error is below $1\%$, which is tolerable in most cases.

\ifx\onecol\undefined
\begin{figure}
	\centering 
	\includegraphics[width=0.5\textwidth]{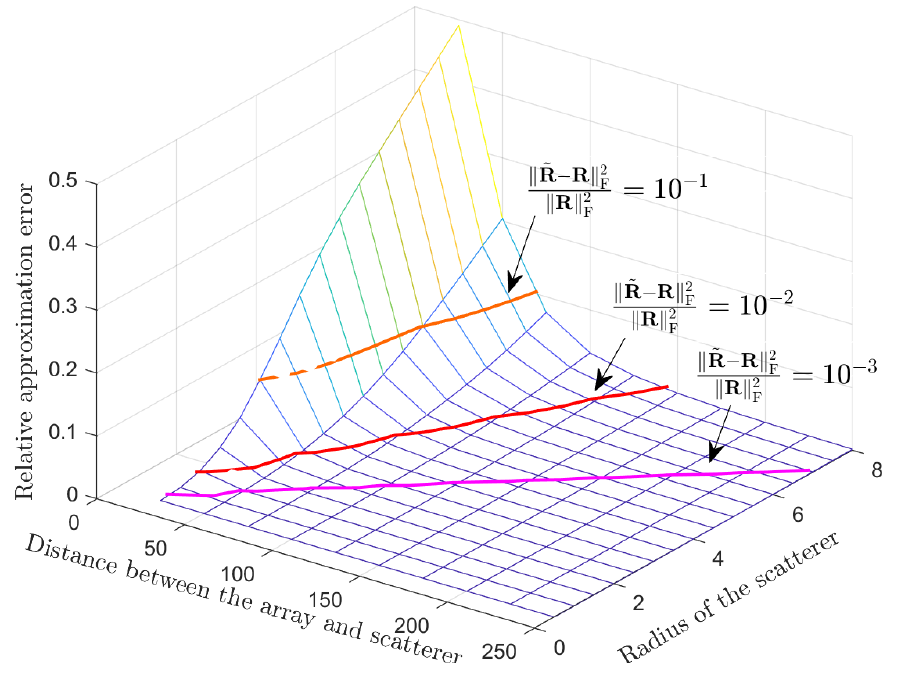} 
	\caption{The correlation function plotted from the approximated analytical expression.} 
	\label{fig_approximation_error}
\end{figure}
\else
\begin{figure}
	\centering 
	\includegraphics[width=0.7\textwidth]{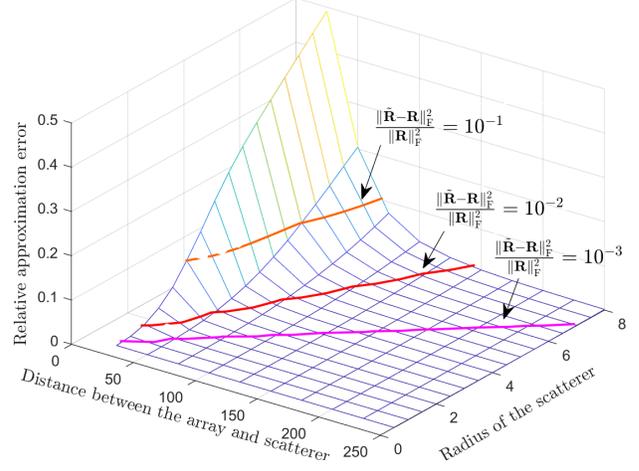} 
	\caption{The correlation function plotted from the approximated analytical expression.} 
	\label{fig_approximation_error}
\end{figure}
\fi

\section{Characteristics of the Proposed Channel Model}
\label{sec_charac}
In this section, we will analyze and show the characteristics of the derived channel model, which can reveal how the scattering environment affects the system performance. 

\subsection{One realization of the random field}

For the derived correlation function $R({\bf r},{\bf r}')$, we have the following expansion $R = \sum_{i=1}^{\infty}\lambda_i \phi_{i}({\bf r})\phi_{i}^{*}({\bf r}')$ from Mercer's theorem, where $\phi({\bf r})$ is the solution of the following integral equation
\begin{equation}
	\begin{aligned}
		\lambda_i \phi_{i}({\bf r})  = \int_V R({\bf r},{\bf r}') \phi_{i}({\bf r}) {\rm d}{\bf r},
	\end{aligned}
\end{equation}
according to \cite{mercer1909xvi}. Then the received field can be constructed by its  Karhunen-Lo{\`e}ve expansion 
\begin{equation}
	\begin{aligned}
		E({\bf r})  = \sum_{i=1}^{\infty} \xi_{i} \phi_{i}({\bf r}),
	\end{aligned}
\end{equation}
where $\lambda_i = {\mathbb E}[\xi_i\xi_i^{*}]$. For a noisy received field $Y({\bf r}) = E({\bf r})+N({\bf r})$ where $R_N({\bf r},{\bf r}') = \sigma^2\delta({\bf r}-{\bf r}')$, the information that can be obtained from the received field is $I(E;Y) = \sum_i{\rm log}(1+\frac{\lambda_i}{\sigma^2})$ \cite{wan2022mutual}.

If we consider discrete samples of the continuous fields, for a $N_y \times N_z$ array at the receiver, we can construct a correlation matrix ${\bf R} \in {\mathbb C}^{(N_y N_z)\times (N_y N_z)}$, where ${\bf R}_{i,j} = R({\bf r}_i,{\bf r}_j)$, ${\bf r}_i = \left[0, \lfloor \frac{i-1}{N_z} \rfloor-\frac{N_y-1}{2}, {\rm mod}\left( i-1,N_z \right)- \frac{N_z-1}{2} \right] $, and ${\bf r}_j = \left[0, \lfloor \frac{j-1}{N_z} \rfloor-\frac{N_y-1}{2}, {\rm mod}\left( j-1,N_z \right)- \frac{N_z-1}{2} \right] $.
From the correlation function of the received field, we can generate the channel by ${\bf h} = {\bf L}{\bf N}$, where ${\bf L}$ is the Cholesky decomposition of the correlation matrix ${\bf R}$, and ${\bf N} \sim \mathcal{CN}(0,{\bf I})$.

\subsection{Fitness to the statistics of practical model}
\begin{figure}[!t]
	\setlength{\abovecaptionskip}{-0.0cm}
	\setlength{\belowcaptionskip}{-0.0cm}
	\centering
	\subfigcapskip -0.5em
	\subfigure[CDL-D model]{
		\includegraphics[width=1.5in]{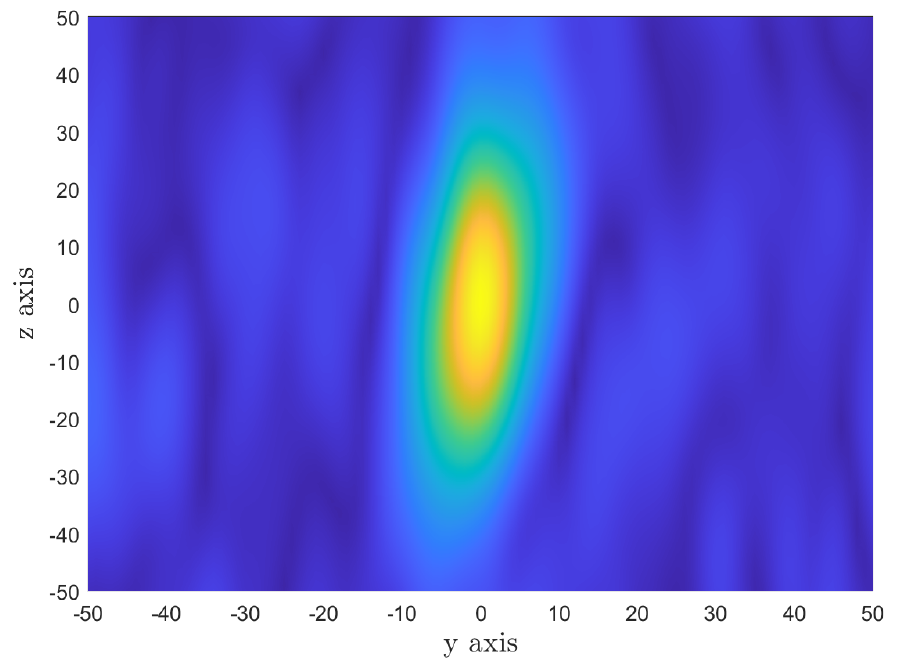}
	}
	\subfigure[proposed coupling model]{
		\includegraphics[width=1.5in]{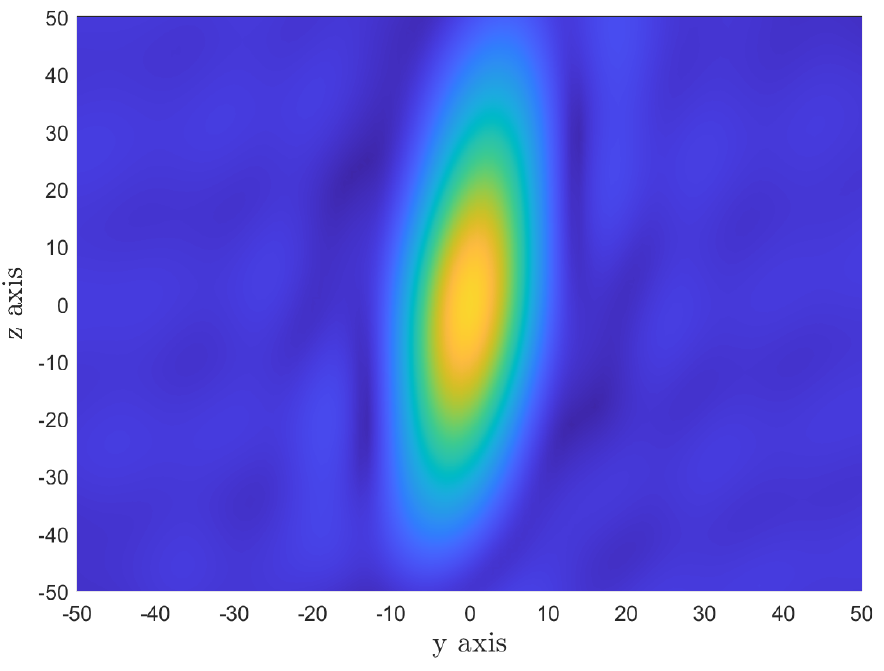}
	}
	
	\subfigure[CDL-A model]{
		\includegraphics[width=1.5in]{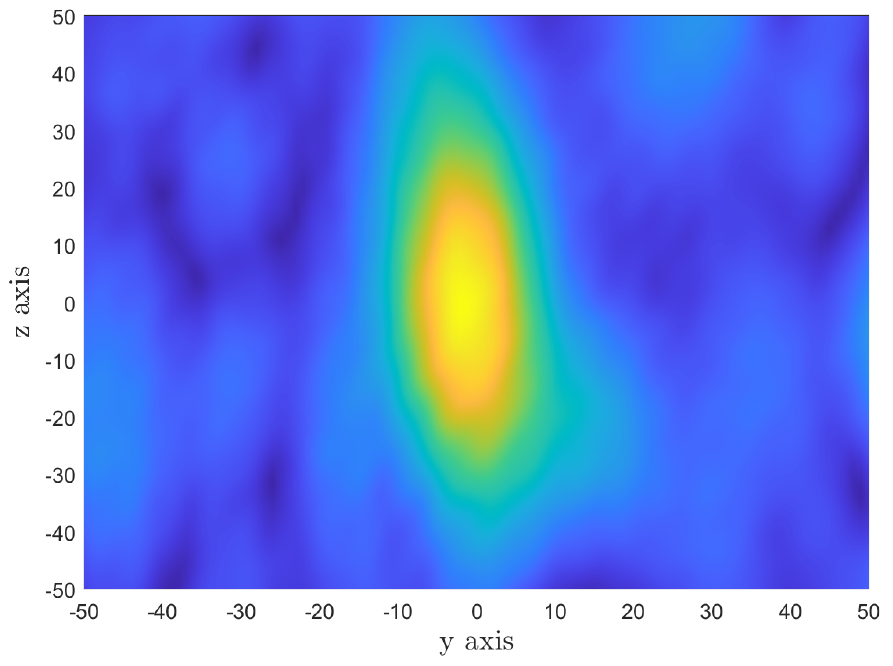}
	}
	\subfigure[proposed coupling model]{
		\includegraphics[width=1.5in]{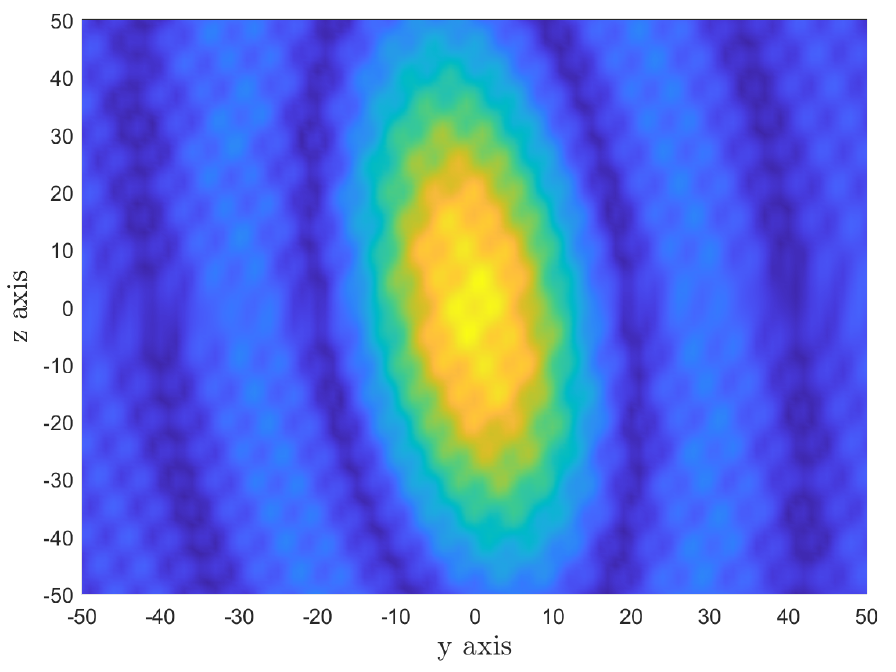}
	}
	\caption{
		Comparison between the field correlation of CDL-A, CDL-D and the proposed coupling model.
	}
	\label{fig_cdl_fit}
\end{figure}

\begin{figure}
	\centering 
	\includegraphics[width=0.5\textwidth]{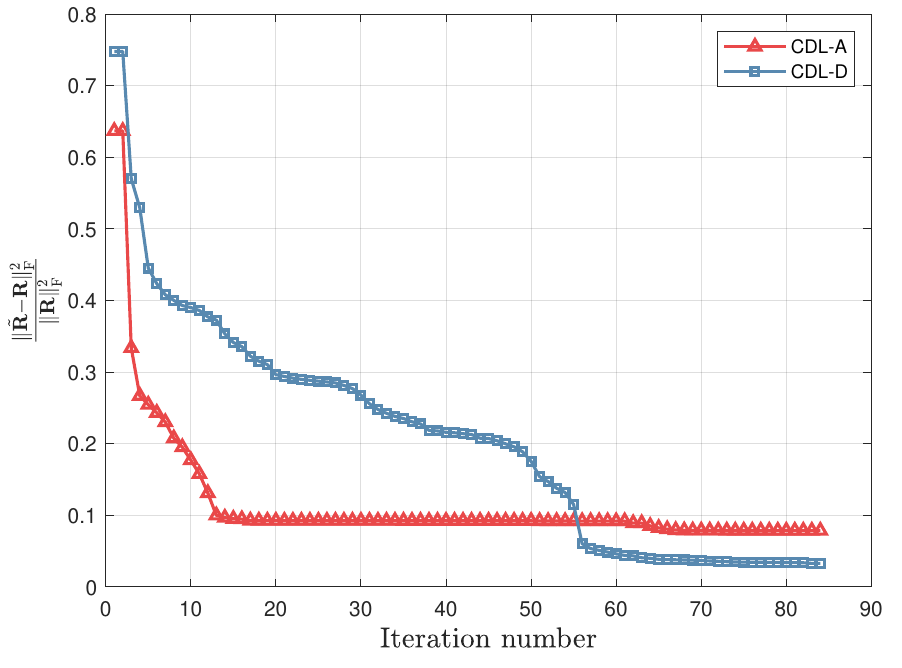} 
	\caption{{The loss function degrades when iteration number increases.}} 
	\label{fig_optimization}
\end{figure}

\begin{figure}
	\centering 
	\includegraphics[width=0.5\textwidth]{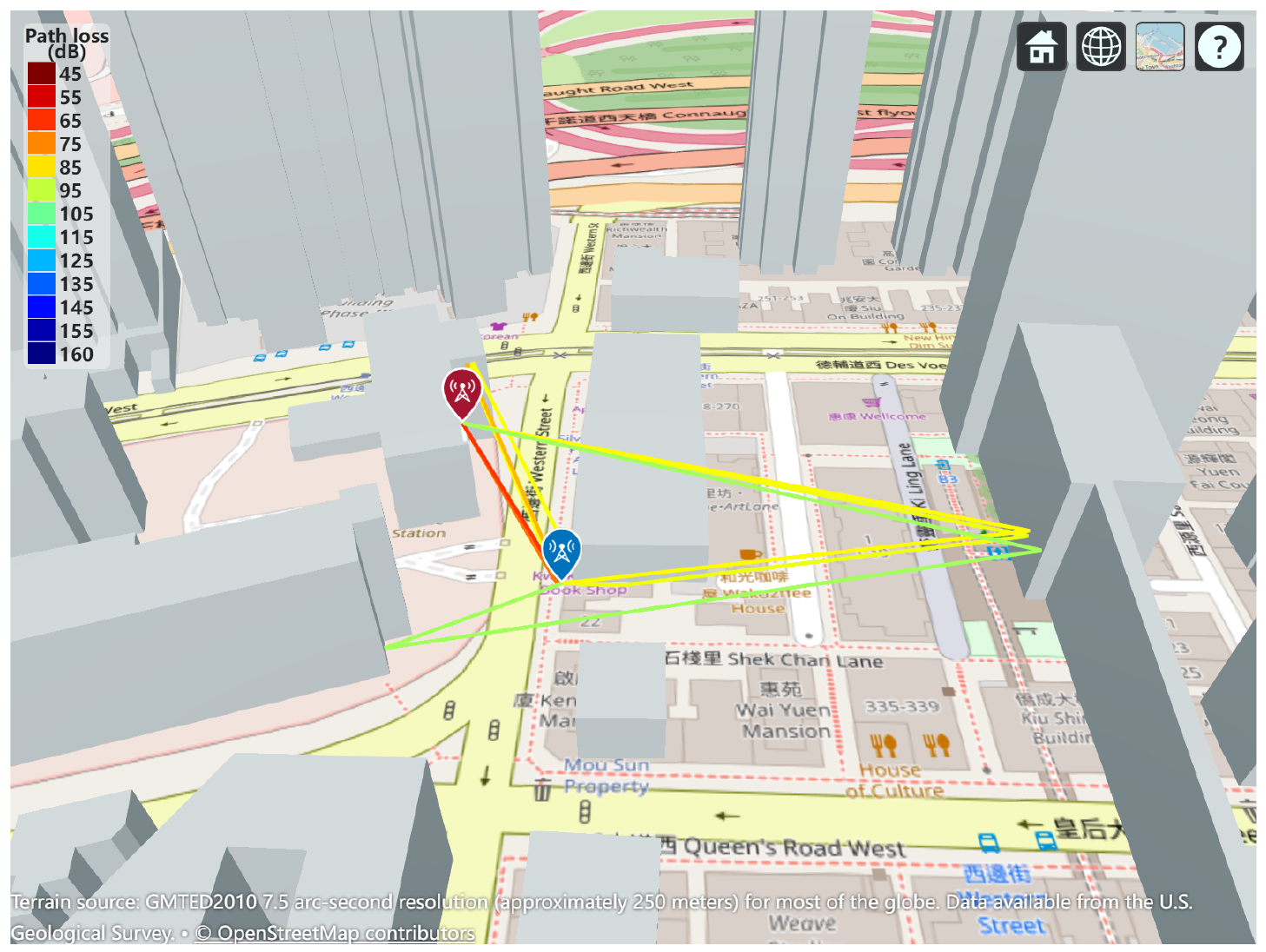} 
	\caption{The model built from ray tracing scheme in Matlab, which is based on buildings in Hong Kong \cite{schaubach1992ray}.} 
	\label{fig_ray_tracing}
\end{figure}

\begin{figure}
	\centering 
	\includegraphics[width=0.5\textwidth]{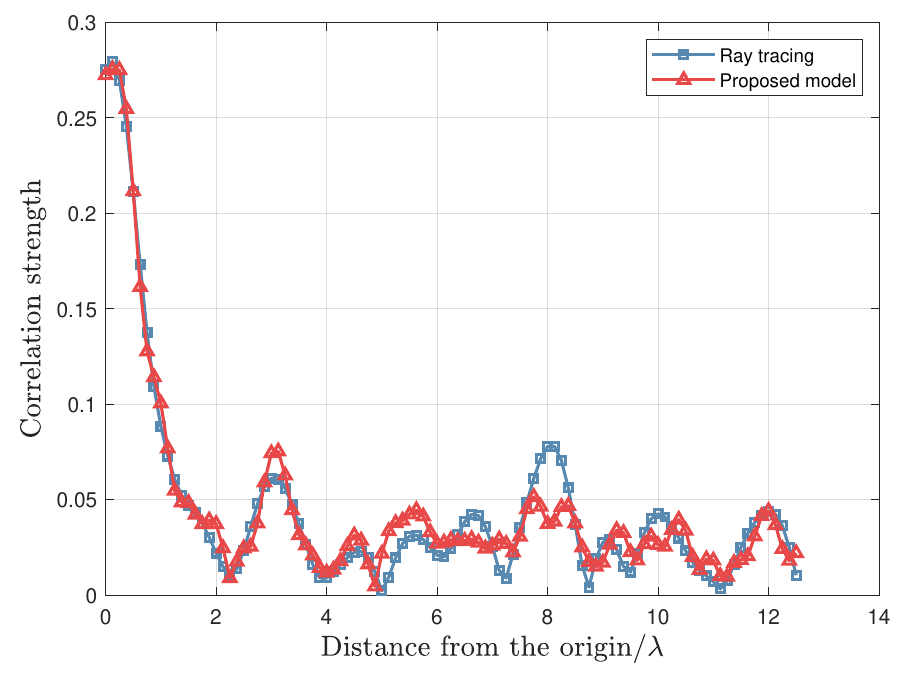} 
	\caption{Comparison between the proposed model and the model built from ray tracing scheme.} 
	\label{fig_ray_tracing_fit}
\end{figure}

In this part we will show the fitness of the proposed model to the statistics of standard 3GPP TR 38.901 CDL model \cite{CDL}. Since CDL model is now widely used in 5G new radio (5G NR) scenarios, the rationality of the proposed analytical correlation function of the channel model can be verified if it can well fit the statistics of the CDL model. 
We simulate the field correlation of CDL-A and CDL-D model, which represent strong scattering and weak scattering scenarios separately. {For the antenna array, we adopt $101\times 101$ array with $\lambda/8$ antenna spacing. We use the proposed analytical model with 3 scatterers to fit the field correlation of the CDL models, which is shown in Fig. \ref{fig_cdl_fit}. We introduce the metric $f = \frac{\| \tilde{\bf R}-{\bf R}_{\rm CDL} \|_{\rm F}^2}{\| {\bf R}_{\rm CDL} \|_{\rm F}^2}$ to depict the difference between the CDL model and the proposed model, and use it as the loss function to optimize the parameters of the proposed model. Specifically, we adopt the quasi-Newton algorithm, where the iteration scheme is 
\begin{subequations}
	\begin{align}
	&{\bf x}_{k+1} = {\bf x}_k - \alpha_k {\bf H}_k \nabla f({\bf x}_k),\\
	& {\bf q}_k = \nabla f({\bf x}_{k+1})-\nabla f({\bf x}_{k}),\\
	& {\bf V}_k = {\bf I}-\frac{{\bf q}_k({\bf x}_{k+1}-{\bf x}_k)^{\rm T}}{{\bf q}_k^{\rm T}({\bf x}_{k+1}-{\bf x}_k)},\\
	&{\bf H}_{k+1} = {\bf V}_k{\bf H}_k {\bf V}_k^{\rm T}+\frac{({\bf x}_{k+1}-{\bf x}_k)({\bf x}_{k+1}-{\bf x}_k)^{\rm T}}{{\bf q}_k^{\rm T}({\bf x}_{k+1}-{\bf x}_k)} .
	\end{align}
\end{subequations}

It is shown in Fig. \ref{fig_cdl_fit} that the proposed model can fit the statistical characteristics of CDL models with few parameters, which verifies its correctness and generalization capability. Then, we show the optimization procedure in Fig. \ref{fig_optimization}, where tolerable loss is achieved by 13 iterations under CDL-A channel model and 56 iterations under CDL-D channel model. 

Furthermore, we fit the proposed model to the model generated by ray tracing scheme to show its fitness to practical scenarios. The transceivers locate in Hong Kong, and the paths between the transceivers are characterized by ray tracing scheme, as shown in Fig. \ref{fig_ray_tracing}. We use quasi-Newton algorithm to fit the proposed model to the model generated by ray tracing scheme. The result is shown in Fig. \ref{fig_ray_tracing_fit}. We can observe that the proposed model can well rebuild the channel with limited parameters.}

{ The benefits of the proposed model is that it is analytical and can be used to obtain the field correlation between any two positions by direct and quick calculation, while the existing models need a large amount of parameters.
For example, the widely-accepted CDL-A model has 23 scatterer clusters and 20 rays in each cluster. Therefore, it is accurate but very complex, making it difficult for further analysis. Moreover, the proposed model provides a correlation function expression of the received field. Therefore, it can be used in channel estimation process to provide prior information for channel estimator.}

\subsection{Impact of the scatterer size on the channel}
{In this part, we will discuss the impact of the scatterer size, how it affects the channel in the wavenumber domain, and when it can not be neglected.}
	
	First note that in the scenario with far field approximation the correlation function of the random channel can be simplified to 
	\begin{equation}
		\begin{aligned}
			R({\bf r}_1,{\bf r}_2) &= \beta \int_V \frac{e^{{\rm j}k\| {\bf r}_1-{\bf r}' \|}}{4\pi \|{\bf r}_1-{\bf r}' \|}\frac{e^{-{\rm j}k\| {\bf r}_2-{\bf r}' \|}}{4\pi \|{\bf r}_2-{\bf r}' \|} f({\bf r}'){\rm d}{\bf r}'
			\\& \overset{r_m \rightarrow 0}{\approx} \beta \int_{V} \frac{e^{{\rm j}k \left( r' - \hat{\bf r}'\cdot {\bf r}_1\right)}e^{-{\rm j}k \left( r' - \hat{\bf r}'\cdot {\bf r}_2\right)}}{16\pi^2 \| {\bf r}_0-{\bf r}' \|^2} f({\bf r}'){\rm d}{\bf r}'
			\\& {\overset{r_s \rightarrow 0}{\approx} \frac{\beta}{16\pi^2 \| {\bf r}_0-{\bf r}'_0 \|^2} e^{-{\rm j}k \hat{\bf r}'_0 \cdot ({\bf r}_1-{\bf r}_2)},}
		\end{aligned}
		\label{equ_far_approx}
	\end{equation}
	where ${\bf r}_0$ is the position of the center of the receiver array, {and ${\bf r}'_0$ is the position of the center of the scatterer. The last approximation is based on the fact that for $r_s = \frac{1}{2n}$, where $n\in \mathcal{Z}^+$, $\int_V f({\bf r}'){\rm d}{\bf r}' = 1$. Moreover, when $n\rightarrow +\infty$, $f({\bf r}')=0$ for any ${\bf r}'\neq{\bf r}'_0$. Therefore, $f({\bf r}')$ approaches $\delta({\bf r}'-{\bf r}'_0)$ when $r_s$ approaches 0}. Therefore, the received field under such approximation is a stationary field, which implies that its correlation function only relies on the distance vector between the two points. If we perform Fourier transformation on the correlation function, we will obtain its power spectrum in the wavenumber domain. To be more specific, we have
	\ifx\onecol\undefined
	\begin{equation}
		\begin{aligned}
			&~~~~\int_{-\infty}^{+\infty}\int_{-\infty}^{+\infty} R(\Delta {\bf r}) e^{{\rm j}(k_y y+ k_z z)} {\rm d}y{\rm d}z \\&= \int_{-\infty}^{+\infty}\int_{-\infty}^{+\infty} \beta_0 e^{-{\rm j}(\hat{\bf r}_x' x+\hat{\bf r}_y' y+\hat{\bf r}_z' z)} e^{{\rm j}(k_y y+ k_z z)} {\rm d}y
			{\rm d}z  \\& = \beta_0 e^{-{\rm j} \hat{\bf r}_x' x} \delta(k_y-\hat{\bf r}_y')\delta(k_z-\hat{\bf r}_z').
		\end{aligned}
		\label{equ_double_delta}
	\end{equation}
	\else
	\begin{equation}
		\begin{aligned}
			&~~~~\int_{-\infty}^{+\infty}\int_{-\infty}^{+\infty} R(\Delta {\bf r}) e^{{\rm j}(k_y y+ k_z z)} {\rm d}y{\rm d}z \\&= \int_{-\infty}^{+\infty}\int_{-\infty}^{+\infty} \beta_0 e^{-{\rm j}(\hat{\bf r}_x' x+\hat{\bf r}_y' y+\hat{\bf r}_z' z)} e^{{\rm j}(k_y y+ k_z z)} {\rm d}y
			{\rm d}z  \\& = \beta_0 e^{-{\rm j} \hat{\bf r}_x' x} \delta(k_y-\hat{\bf r}_y')\delta(k_z-\hat{\bf r}_z').
		\end{aligned}
	\end{equation}
	\fi
	Therefore, the Fourier transform of the far-field correlation function reveals the angular concentration of the scattering regions in the wavenumber domain. A scattering region with the azimuth angle $\theta$ and elevation angle $\phi$ will lead to a single point $[\cos(\theta)\sin(\phi),\sin(\theta)\sin(\phi),\cos(\phi)]$ in the wavenumber domain of the received field under far-field assumption. If we sample the continuous received electromagnetic fields to obtain a correlation matrix ${\bf R} \in \mathbb{C}^{N_y \times N_z}$ where ${\bf R}_{i,j} = R({\bf r}_0,{\bf r})$ and ${\bf r} = [0,\pm n_yd_y,\pm n_zd_z]$, we can use Fourier transform matrices ${\bf F}_1$ and ${\bf F}_2$ instead continuous Fourier transform to find the angular sparsity of the correlation matrix by ${\bf F}_1{\bf R}{\bf F}_2^{\rm H}$. Specifically, the Fourier transform matrix ${\bf F}_1$ and ${\bf F}_2$ can be constructed by ${\bf F}_{1,i,j} = e^{{\rm j}\frac{2k}{N_y-1}(i-\frac{N_y+1}{2})(j-\frac{N_y+1}{2})d_y}$ and ${\bf F}_{2,i,j} = e^{{\rm j}\frac{2k}{N_z-1}(i-\frac{N_z+1}{2})(j-\frac{N_z+1}{2})d_z}$.
	
	{If the channel is sparse, from the law of large numbers, we know that $h$ has power peaks in wavenumber domain,
	which can be used in channel estimation of reconstruction procedure to improve the accuracy.} Specifically, if we reshape the vector ${\bf h}$ to a matrix ${\bf H}_{i,:} = {\bf h}_{(i-1)*N_z+1:i*N_z,1}^{\rm T}$, its sparsity in wavenumber domain can be expressed as follows:
	\begin{equation}
		\begin{aligned}
			&{\mathbb E}[\left|({\bf F}_1{\bf H}{\bf F}_2)_{i,j}\right|^2] = {\mathbb E}\left[ \left| \sum_{i'}\sum_{j'}{\bf F}_{1,i,j'}{\bf H}_{i',j'}{\bf F}_{2,i',j}^{\rm H} \right|^2 \right]
			\\& ={\mathbb E}\left[ \sum_{i'_1,i'_2,j'_1,j'_2}{\bf F}_{1,i,j'_1}{\bf H}_{i'_1,j'_1}{\bf F}_{2,i'_1,j}^{\rm H}{\bf F}^{*}_{1,i,j'_2}{\bf H}^{*}_{i'_2,j'_2}{\bf F}_{2,i'_2,j}^{\rm T}\right]
			\\& = \sum_{i_1',i_2',j'_1,j'_2} e^{{\rm j}\frac{2k}{N_y-1}(i-\frac{N_y+1}{2})(j_1'-j_2')d_y}R({\bf r}_1,{\bf r}_2)
			\\& ~~~~e^{{\rm j}\frac{2k}{N_z-1}(j-\frac{N_z+1}{2})(i_2'-i_1')d_z},
		\end{aligned}
	\end{equation}
	where ${\bf r}_1 = [0,(i_1'-\frac{N_y+1}{2})d_y,(j_1'-\frac{N_z+1}{2})d_z]$ and ${\bf r}_2 = [0,(i_2'-\frac{N_y+1}{2})d_y,(j_2'-\frac{N_z+1}{2})d_z]$ respectively. From (\ref{equ_far_approx}) it is easy to know that when $r_m$ and $r_s$ approximates 0, ${\mathbb E}[\left|({\bf F}_1{\bf H}{\bf F}_2)_{i,j}\right|^2]$ will reaches a peak value compared to its neighbors, which is in the form of products of sinc function as the discretized form of (\ref{equ_double_delta}). 
	
\begin{figure}
		\centering 
		\includegraphics[width=0.5\textwidth]{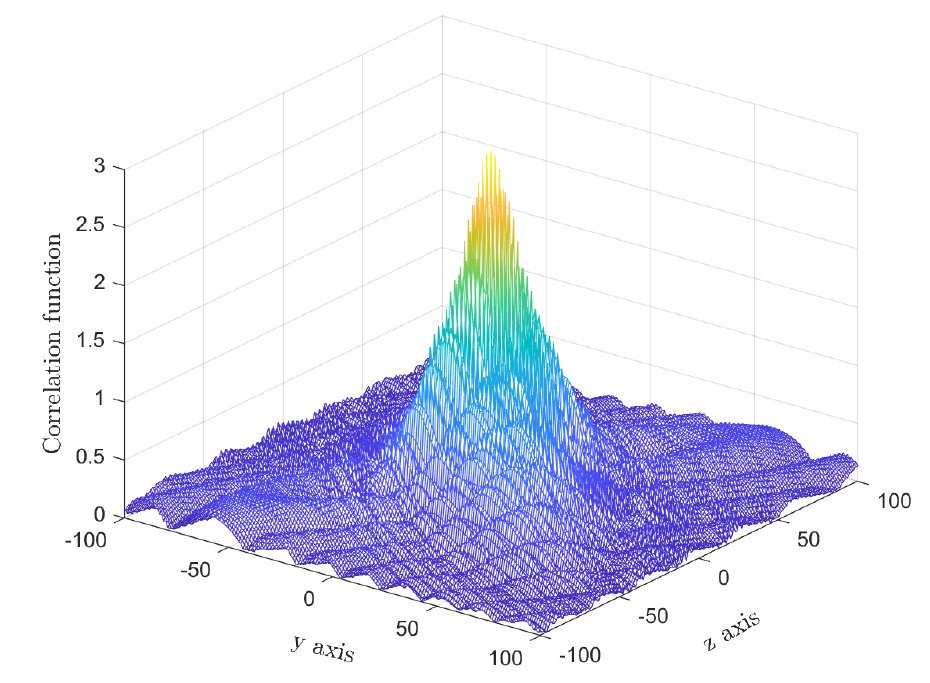} 
		\caption{The correlation function of the received field. } 
		\label{fig_Correlation_Fourier}
	\end{figure}

\begin{figure}
		\centering 
		\includegraphics[width=0.5\textwidth]{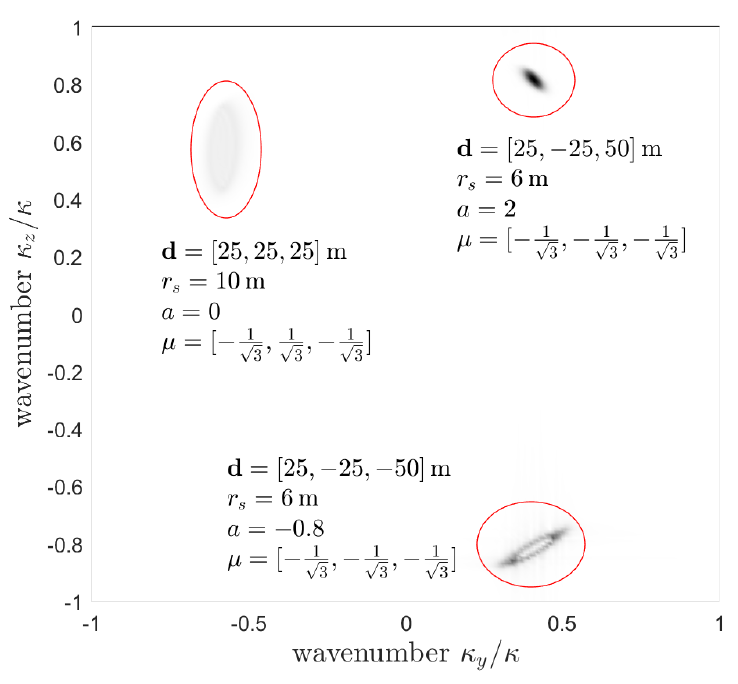} 
		\caption{The Fourier transform of the correlation function of the received field. Three spots in the figure corresponds to three scattering regions in the space.} 
		\label{fig_Fourier}
	\end{figure}

	For the near-field scattering scenario, the scattering region will correspond to an area rather than a point in the wavenumber domain. The shape of the area reflects the size, directions and concentration parameters of the scattering region. { It is also worth noting that the area relates to the concept called spatial bandwidth \cite{bucci1987spatial}. The larger the area is, the larger the spatial bandwidth is, which provides more possible DoF for the wireless communication system. Further discussion about the DoF will be presented in the following subsection.} We plot the correlation function of a generated channel in Fig. \ref{fig_Correlation_Fourier}, and its Fourier transform in Fig. \ref{fig_Fourier}. Three scattering regions are located in the space, with coordinates ${\bf d}=[25,25,25]\,{\rm m}$, ${\bf d}=[25,-25,50]\,{\rm m}$, and ${\bf d}=[25,-25,-50]\,{\rm m}$ separately. While little information can be directly observed from the figure of the correlation function of the received field, the Fourier transform of the correlation function reflects its sparsity in the wavenumber domain. Three shaded areas in Fig. \ref{fig_Fourier} correspond to three scattering regions in the settings, with their respective parameters labeled adjacent to the shaded areas. It is shown that when $R$ increases and $r$ decreases, the size of shaded areas will increase, which corresponds to larger angular expansion in the wavenumber domain. When $a$ tends to infinity, the shaded area tends to a single point. When $a$ tends to $-1$, the shaded area tends to a circle, which aligns with the definition of function $f({\bf r})$.    
	

{ Then we will provide quantitative analysis to show how $r_s$ influences the accuracy of the model and when it can not be ignored.}
It is well known that the Rayleigh distance, also called as Fraunhofer distance, is $d = \frac{8r_{\rm m}^2}{\lambda}$, where $d$ is the distance from the antenna array, and $r_{\rm m} = {\rm max}\| {\bf r}\|$ is the radius of the antenna array \cite{selvan2017fraunhofer}. The Rayleigh distance is defined by the distance where $\frac{\pi}{8}$ phase error is observed on the antenna array. If we further consider the size of the scatterer, we have the channel response as $h({\bf r}',{\bf r}) = e^{{\rm j}\frac{2\pi}{\lambda}\|{\bf r}'-{\bf r}\|}$, where ${\bf r}' = {\bf d}+\boldsymbol{\rho}$ is the position of one point on the scatterer. In {\bf Lemma 2} we extend the Rayleigh distance considering scatterers
\begin{lemma}[Extension of Rayleigh distance considering scatterer size]
	The size of scatterer can be neglected when $r_{\rm s} \leqslant \frac{\lambda}{16}$ and {$d \leqslant \frac{8(r_{\rm s}+r_{\rm m})^2}{\lambda-16r_s}$}. Otherwise, the scatterer size should be considered in the channel model. Under this scenario, when {$d \leqslant \frac{8(r_{\rm s}+r_{\rm m})^2}{\lambda} $}, the scatterer and the antenna array are in the near-field region. When {$d \geqslant \frac{8(r_{\rm s}+r_{\rm m})^2}{\lambda}$}, the scatterer and the antenna array are in the far-field region. 
\end{lemma}
\begin{IEEEproof}
	See Appendix B.
\end{IEEEproof}

From {\bf Lemma 2} we know that unless the scatterer is small enough (for frequency of $1\,{\rm GHz}$ the radius of the scatterer should be smaller than 0.0187\,{\rm m}), neglecting the size of the scatterer and simply view it as a point will be inaccurate. Therefore, considering parameters of scatterer is of necessity in channel modeling especially in near-field communication scenarios.

\subsection{Channel DoF of the proposed model}

In this subsection we will discuss how the parameters influence the performance of the system from the degree of freedom (DoF) perspective.
The DoF of the channel depends on the eigenvalue distribution of the model. If the eigenvalue decay rate is slow, there exist multiple subchannels that can support communication at a certain rate, leading to greater DoF. On the contrary, if few eigenvalues are obviously larger than other eigenvalues, the DoF will be small \cite{bjornson2020rayleigh}. 

\begin{figure}[t]
	\centering 
	\includegraphics[width=0.4\textwidth]{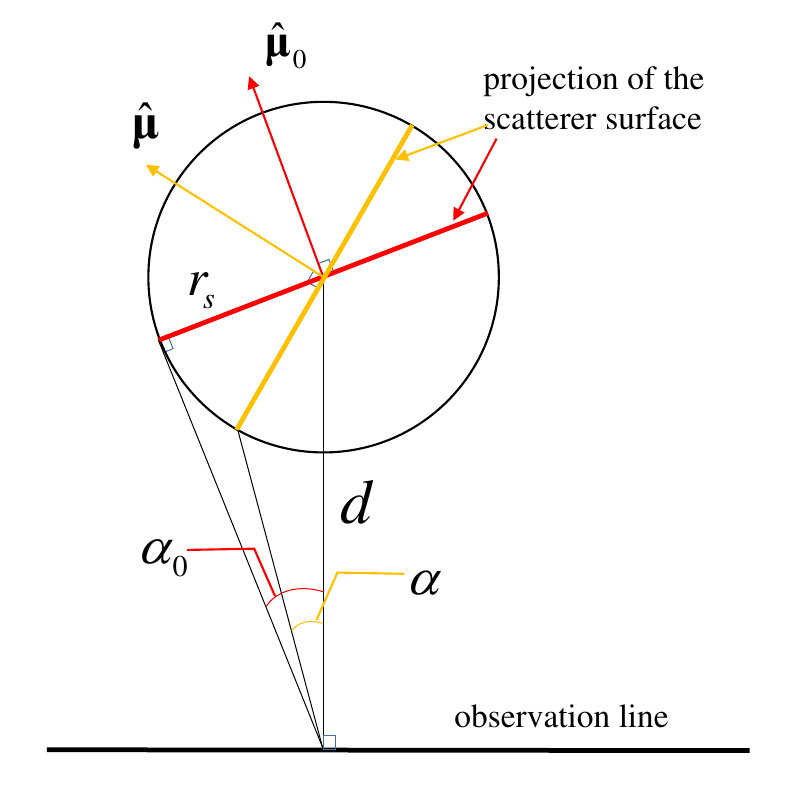} 
	\caption{{Concerning the spatial bandwidth with respect to different $\boldsymbol{\hat{\mu}}$.}} 
	\label{fig_mu}
\end{figure}

{ {We will first provide some insights of the DoF from the spatial bandwidth \cite{bucci1987spatial} perspective and then verify them by numerical analysis of the proposed model. The spatial bandwidth characterizes the band-limiting effect of electromagnetic fields in the wavenumber domain, which is similar to the classical bandwidth that depicts a function's band-limiting effect in the frequency domain. The spatial bandwidth shows the electromagnetic fields' DoF through spatial sampling. In \cite{bucci1987spatial}, the scattered electromagnetic waves ${\bf E}({\bf r})$ are observed on an infinite line or region at the receiver. In this paper we adopt scalar form of the electromagnetic field, leading to $E({\bf r}) = \int_{V} g({\bf r},{\bf r}')(k^2({\bf r}')-k_0^2)  E({\bf r}'){\rm d}{\bf r}'$. We have $\bar{E}({\bf r}) = E({\bf r})e^{-{\rm j}k_0\left\|{\bf r} \right\|}$ to single out the phase factor introduced by the distance, and $X({\bf r}) = (k^2({\bf r})-k_0^2)  E({\bf r})$ on the scatterer surface. Then we have
\begin{equation}
	\bar{E}({\bf r}) = \int_V \bar{g}({\bf r},{\bf r}') X({\bf r}') {\rm d}{\bf r}',
\end{equation}
where $\bar{g}({\bf r},{\bf r}') = \frac{1}{2\pi}\frac{e^{{\rm j}k_0(\left\| {\bf r}-{\bf r}' \right\|-\left\| {\bf r} \right\|)}}{\left\| {\bf r}-{\bf r}' \right\|}$. For simplicity we focus on the one-dimensional receiver and abbreviate $\bar{E}({\bf r})$ as $\bar{E}({r})$, $\bar{g}({\bf r},{\bf r}')$ as $\bar{g}({r},{\bf r}')$, where $r$ is along a chosen line determined by ${\bf r}$ in the three-dimensional space. To show the how $\bar{E}({r})$ is band-limited in the wavenumber domain, we introduce $\bar{E}_w({r}) = \bar{E}({r}) * \frac{\sin w r}{r}$ which performs low-pass filtering on $\mathscr{F}[\bar{E}({r})]$. Then we have 
\begin{equation}
	\bar{E}_w({r}) = \int_V \bar{g}_w({r},{\bf r}') X({\bf r}') {\rm d}{\bf r}',
\end{equation}
where 
\begin{equation}
	\begin{aligned}
	\bar{g}_w({r},{\bf r}') &= \frac{1}{2\pi^2} \int_{-\infty}^{+\infty} \frac{\sin w(r-\xi)}{r-\xi}\frac{e^{{\rm j}k_0(\left\| \boldsymbol{\xi}-{\bf r}' \right\|-\left\| \boldsymbol{\xi} \right\|)}}{\left\| \boldsymbol{\xi}-{\bf r}' \right\|}{\rm d}\xi
	\\& \overset{a}{=} \frac{1}{2\pi^2{\rm j}} \int_{C^+} \frac{e^{ {\rm j}w(r-\xi)}}{r-\xi}\frac{e^{{\rm j}k_0(\left\| \boldsymbol{\xi}-{\bf r}' \right\|-\left\| \boldsymbol{\xi} \right\|)}}{\left\| \boldsymbol{\xi}-{\bf r}' \right\|}{\rm d}\xi 
	\\& ~~~~- \frac{1}{2\pi^2{\rm j}} \int_{C^-} \frac{e^{ -{\rm j}w(r-\xi)}}{r-\xi}\frac{e^{{\rm j}k_0(\left\| \boldsymbol{\xi}-{\bf r}' \right\|-\left\| \boldsymbol{\xi} \right\|)}}{\left\| \boldsymbol{\xi}-{\bf r}' \right\|}{\rm d}\xi 
	\\&~~~~+ \bar{g}({r},{\bf r}'),
	\end{aligned}
\end{equation}
in which $\overset{a}{=}$ is from the residual theorem, $C^+$ and $C^-$ are two paths above and below the real axis.
The spatial bandwidth is the minimum $w$ that makes $\left\|\bar{E}(r)-\bar{E}_w(r)\right\|$ small enough. 
We have 
\begin{equation}
	\begin{aligned}
	\left\|\bar{E}(r)-\bar{E}_w(r)\right\| &=  \left[\int_{-\infty}^{+\infty}  \left|\bar{E}(r)-\bar{E}_w(r)\right|^2{\rm d}r \right]^{\frac{1}{2}}
	\\& \leqslant \underset{{\bf r}'}{\rm max} \left[ \int_{-\infty}^{+\infty} \left| \Delta \bar{g}({r},{\bf r}') \right|^2{\rm d}{\bf r}'  \right]^{\frac{1}{2}} 
	\\& ~~~~\cdot \int_V \left| X({\bf r}') \right| {\rm d}{\bf r}',
	\end{aligned}
\end{equation}
where $\Delta \bar{g} = \bar{g}-\bar{g}_w$ is the item corresponds to the spatial bandwidth $w$. We can further express it by
\begin{equation}
	\begin{aligned}
		\Delta \bar{g}({r},{\bf r}')& =  \frac{1}{2\pi^2{\rm j}} \int_{C^+} \frac{e^{ {\rm j}w(r-\xi)}}{r-\xi}\frac{e^{{\rm j}k_0(\left\| \boldsymbol{\xi}-{\bf r}' \right\|-\left\| \boldsymbol{\xi} \right\|)}}{\left\| \boldsymbol{\xi}-{\bf r}' \right\|}{\rm d}\xi 
		\\& ~~~~- \frac{1}{2\pi^2{\rm j}} \int_{C^-} \frac{e^{ -{\rm j}w(r-\xi)}}{r-\xi}\frac{e^{{\rm j}k_0(\left\| \boldsymbol{\xi}-{\bf r}' \right\|-\left\| \boldsymbol{\xi} \right\|)}}{\left\| \boldsymbol{\xi}-{\bf r}' \right\|}{\rm d}\xi. 
	\end{aligned}
\end{equation}}
{
The following part is similar to \cite{bucci1987spatial}, which shows that when $w>{\rm max}\frac{\partial \left( k_0\left( \left\| \boldsymbol{\xi}-{\bf r}' \right\|-\left\| \boldsymbol{\xi} \right\| \right) \right)}{\partial \xi}$, $\Delta \bar{g}$ converges to 0 faster than any power. Moreover, when $w<{\rm max}\frac{\partial \left( k_0\left( \left\| \boldsymbol{\xi}-{\bf r}' \right\|-\left\| \boldsymbol{\xi} \right\| \right) \right)}{\partial \xi}$, $\Delta \bar{g}\approx \bar{g}$. Therefore, $w_0 = {\rm max}\frac{\partial \left( k_0\left( \left\| \boldsymbol{\xi}-{\bf r}' \right\|-\left\| \boldsymbol{\xi} \right\| \right) \right)}{\partial \xi}$ can be chosen as the spatial bandwidth of the received electromagnetic field. From geometrical analysis, it is easy to find that the maximum of $\frac{\partial \left( k_0\left( \left\| \boldsymbol{\xi}-{\bf r}' \right\|-\left\| \boldsymbol{\xi} \right\| \right) \right)}{\partial \xi}$ only depends on the radius $r_s$ of the scatterer, and the inner structure of the scatterer does not have obvious influence on the DoF. Moreover, we can bound $w_0$ by $\frac{2\pi r_s}{\lambda}<w_0<\sqrt{2}\frac{2\pi r_s}{\lambda}$, where $r_s$ is the radius of the scatterer. The lower bound of $w_0$ is achieved when $\boldsymbol{\hat{\mu}} = \boldsymbol{\hat{\mu}}_0$ satisfies the condition that the corresponding scatterer surface is tangent to $OF$, as shown in Fig. \ref{fig_mu}. When $\boldsymbol{\hat{\mu}} \neq \boldsymbol{\hat{\mu}}_0$, the lower bound of $w_0$ will be less than $\frac{2\pi r_s}{\lambda}$. Specifically, we can obtain
\begin{equation}
	\begin{aligned}
		w_0&>\frac{2\pi d }{\lambda} 2\sin \frac{\alpha}{2} \cos \frac{\alpha}{2} = \frac{2\pi d \sin \alpha}{\lambda}
		\\& = \frac{2\pi d}{\lambda} \frac{r_s \cos(\alpha_0+\theta)}{\sqrt{(d-r_s\sin(\alpha_0+\theta))^2+(r_s\cos(\alpha_0+\theta))^2}}
		\\& = \frac{2\pi d}{\lambda}\frac{r_s\cos(\alpha_0+\theta)}{\sqrt{d^2+r_s^2-2dr_s\sin(\alpha_0+\theta)}},
	\end{aligned}
\end{equation}
}
{
where $\sin(\alpha_0)=\frac{r_s}{d}$ and $\cos(\theta) = \boldsymbol{\hat{\mu}}\cdot\boldsymbol{\hat{\mu}}_0 $.}
}

{{
From the above analysis we know that the spatial bandwidth and channel DoF mainly rely on the outermost layer of the scattering region. To be more specific, smaller $r_s$ leads to smaller scattering regions, which reduces the DoF. When $a$ is smaller than $0$, the scattering region can be viewed as the outermost circle combined with the inner part, and the DoF will not change heavily with $a$. When all the scattering power comes from the outermost circle, which plays the most important role in affecting the DoF, the DoF will be the largest. Therefore, the DoF reaches the maximum when $a$ approaches $-1$. For two-dimensional receiver adopted in this paper, we can decompose the surface in two different directions, each with spatial bandwidth $w_0$.}
}

{

Then, we plot the eigenvalues of the correlation matrix when the radius and shape of the scattering region vary in Fig. \ref{fig_eigen_a} and Fig. \ref{fig_eigen_r}. We can observe that the DoF of the channel will increase with the radius $r$ of the scatterer, and decrease when $a$ increases. When $a$ approximates $-1$, which corresponds to the case that the scatterer tends to a ring, the DoF of the channel reaches the maximum, which coincides with the spatial bandwidth analysis.

Note that the spatial bandwidth analysis is based on the infinitely large observation region of the received electromagnetic fields. For practical scenarios with limited observation region, the observed field can not be strictly band-limited in the wavenumber domain. Moreover, the spatial sampling period should not be equal to that in sampling theorem because the points far away from the scatterer are not as important as the ones close to the scatterer. This problem is discussed in \cite{wavethoeyofinformation} and the tool of cut-set integral is introduced. Results of the approximated DoF considering a closed surface as the receiver that encloses the source are discussed in \cite{wavethoeyofinformation}, which we will follow to provide DoF bounds in the scenario with square receiving surface.} 

\begin{figure}
	\centering 
	\includegraphics[width=0.5\textwidth]{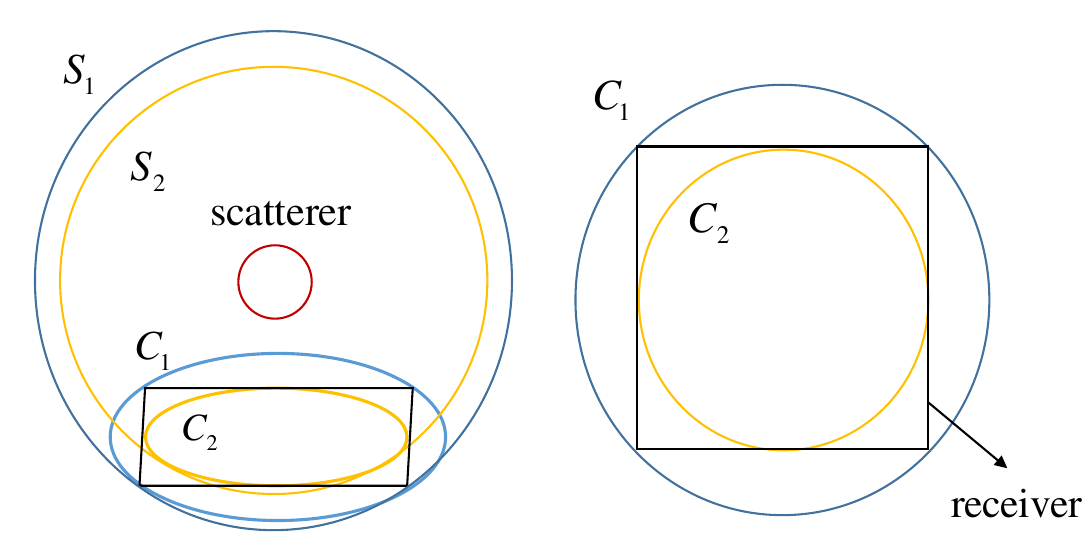} 
	\caption{The receiving surface that is inscribed in $C_1$ and externally-tangent to $C_2$, where $C_1$ and $C_2$ are on $S_1$ and $S_2$ separately.} 
	\label{fig_circles}
\end{figure}

{
Here we discuss a simple scenario that the line between the center of the scatterer and the center of the receiving surface is vertical to the receiving surface. We construct two spheres concentric with the scatterer region. These spheres satisfy the condition that the receiving surface is inscribed in a circle $C_1$ on the large sphere $S_1$, and its four sides are externally-tangent to a circle $C_2$ on the small sphere $S_2$, as shown in Fig. \ref{fig_circles}. We denote the two spherical caps of $S_1$ divided by $C_1$ as $S'_1$ and $S''_1$, where $S'_1$ is the larger one. Similarly we have $S'_2$ and $S''_2$. Since the information flows through any closed surface that enclose the scatterer should be the same, we know that the electromagnetic fields on $C_1$ and $S''_1$ have the same DoF, so as $C_2$ and $S''_2$. According to \cite{wavethoeyofinformation} we know that the DoF on the sphere $S_1$ and $S_2$ are $N_0 = O(r_s^2/\lambda^2)$. From symmetry on the sphere and simple geometry, the DoF $N_1$ on $S''_1$ can be expressed by
\begin{equation}
	N_1 \approx N_0\frac{\mathcal{A}_{S''_1}}{\mathcal{A}_{S_1}}= N_0\frac{2\pi\sqrt{d^2+r_m^2}(\sqrt{d^2+r_m^2}-d)}{4\pi(d^2+r_m^2)}.
\end{equation}
Similarly we know that the DoF $N_1$ on $S''_1$ can be expressed by $N_2 \approx N_0 \frac{2\pi\sqrt{2d^2+r_m^2}(\sqrt{2d^2+r_m^2}-\sqrt{2}d)}{4\pi(2d^2+r_m^2)}$.
Then we have $N_2\leqslant N_{\rm receiver} \leqslant N_1$. Note that when $d\gg r_m$, both $N_1$ and $N_2$ approximates $O(\frac{r_s^2r_m^2}{\lambda^2d^2})$, which coincides with \cite{dardari2020communicating}. On the contrary, if $r_m \gg d$, $N_{\rm receiver} \approx \frac{N_0}{2}$, because it can be viewed as an infinitely-large surface which gets half of the overall electromagnetic waves out of the scatterer. Under this scenario the DoF has little relationship with the distance between the scatterer and the receiver, which coincides with \cite{bucci1987spatial}. Here the asymmetry introduced by $\hat{\boldsymbol{\mu}}$ is not considered since it is hard to evaluate. By considering this asymmetry a more accurate result will be obtained.
}

\ifx\onecol\undefined
\begin{figure}
	\centering 
	\includegraphics[width=0.5\textwidth]{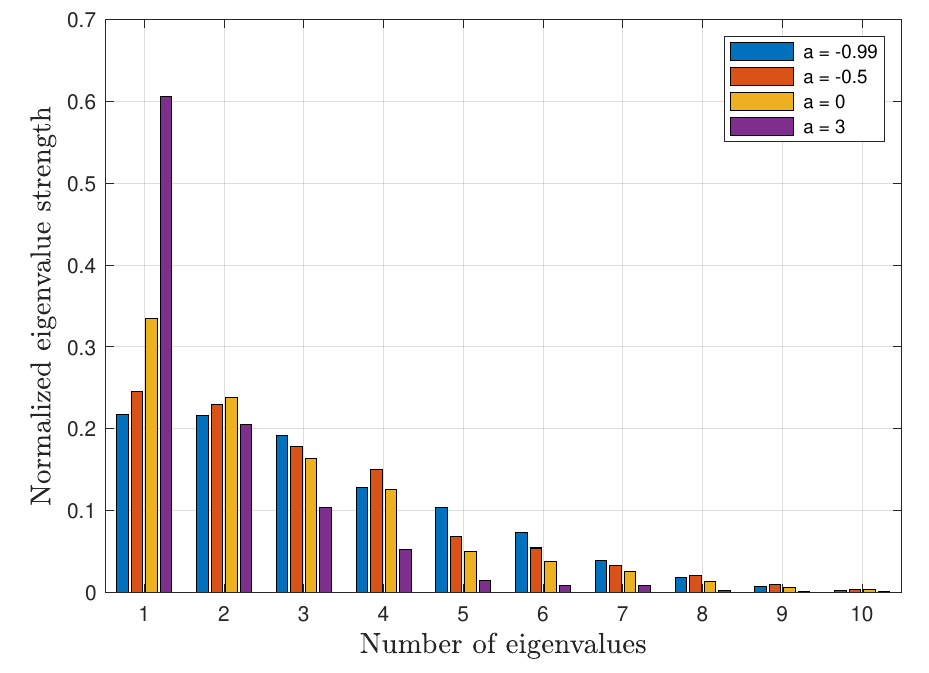} 
	\caption{The eigenvalues of the correlation matrix in decreasing order with ${\bf d}$ fixed to $[-100,100,-100]\,{\rm m}$, $\boldsymbol{\mu}$ fixed to $[-\frac{1}{\sqrt{3}},\frac{1}{\sqrt{3}},-\frac{1}{\sqrt{3}}]$, and $r=5\,{\rm m}$. The concentration parameter $a$ varies.} 
	\label{fig_eigen_a}
\end{figure}
\begin{figure}
	\centering 
	\includegraphics[width=0.5\textwidth]{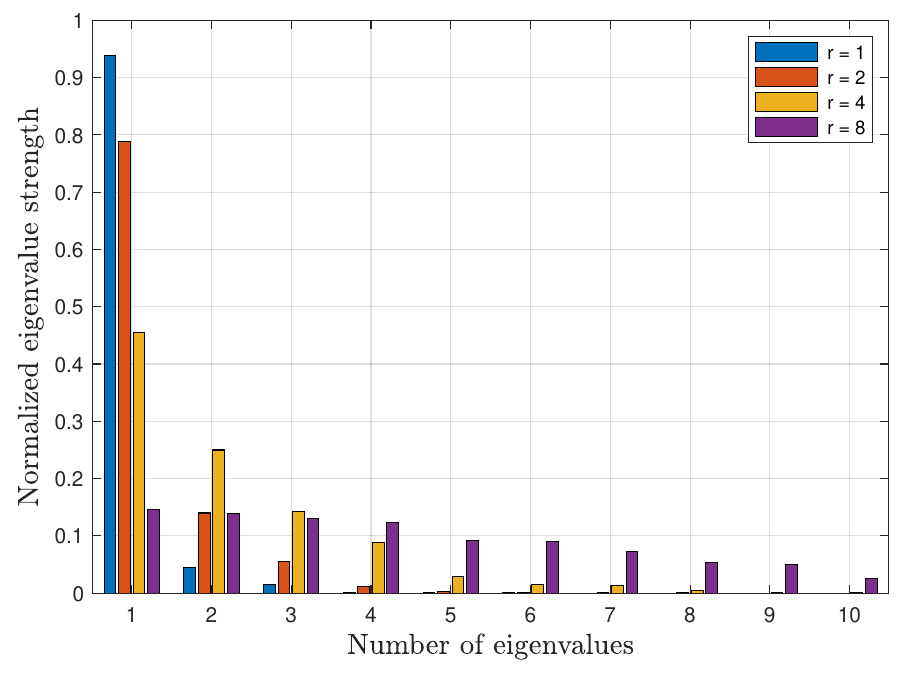} 
	\caption{The eigenvalues of the correlation matrix in decreasing order with ${\bf d}$ fixed to $[-100,100,-100]\,{\rm m}$, $\boldsymbol{\mu}$ fixed to $[-\frac{1}{\sqrt{3}},\frac{1}{\sqrt{3}},-\frac{1}{\sqrt{3}}]$, and $a=0$. The radius $r$ varies.} 
	\label{fig_eigen_r}
\end{figure}
\else
\begin{figure}[!t]
	\setlength{\abovecaptionskip}{-0.0cm}
	\setlength{\belowcaptionskip}{-0.0cm}
	\centering
	\subfigcapskip -1em
	\subfigure[$r=5\,{\rm m}$, the concentration parameter $a$ varies]{
		\includegraphics[width=3.5in]{figs/eigen_a.pdf}
	}
	\subfigure[$a=0$, the radius $r$ varies]{
		\includegraphics[width=3.5in]{figs/eigen_r.pdf}
	}
	\caption{
		The eigenvalues of the correlation matrix in decreasing order with ${\bf R}$ fixed to $[-100,100,-100]\,{\rm m}$, $\boldsymbol{\mu}$ fixed to $[-\frac{1}{\sqrt{3}},\frac{1}{\sqrt{3}},-\frac{1}{\sqrt{3}}]$.
	}
	\label{fig_eigen}
\end{figure}
\fi

\section{Channel Estimation based on the Proposed Model}
\label{Sec_estimation}

After discussing the properties of the analytical channel model, we will propose a near-field channel estimation scheme based on the model. {{We first perform Fourier transformation on the observed field to capture the power peaks in the wavenumber domain. Then we use the }{ proposed model, which provides the prior information of electromagnetic fields, to reconstruct an approximate channel correlation function. This approach is similar to the subspace based channel estimation scheme in \cite{demir2022channel}, which constructs the correlation function based on isotropic scattering field. Compared to \cite{demir2022channel}, our scheme provides more prior information of electromagnetic fields of the fields by using the proposed channel model. Therefore, it can achieve better performance than the existing schemes. }}In the channel estimation procedure, the received field is denoted by ${\bf y}=\sqrt{P}{\bf h}+{\bf n}$, where $P$ is the signal-to-noise ratio, ${\bf h}$ is generated from the channel coupling matrix, and ${\bf n} \sim \mathcal{CN}(0,{\bf I})$ is the noise vector. 

\subsection{LS channel estimation}
The simplest channel estimation scheme is the least square (LS) channel estimation, which leads to $\tilde{\bf h} = {\bf y}/\sqrt{P}$.

\subsection{OMP based channel estimation}

Another widely-used scheme is orthogonal matching pursuit (OMP) \cite{wang2012generalized}, which performs well when the signal is sparse. For the transform matrix ${\bf W}$ in the three-dimensional domain, we adopt the codebook in \cite{wu2023multiple}. Since the three-dimensional codebook provide approximate orthogonal basis for the near-field channel, it fully exploits the angular and distance information of the channel. We have ${\bf y} = {\bf W}{\bf h}_m+{\bf n}$, and the three-dimensional near-field (TDN) OMP algorithm for the channel estimation problem is shown in Algorithm \ref{alg:TDN}.

\begin{algorithm}[t]
	\caption{TDN OMP} 
	\label{alg:TDN}
	\hspace*{0.02in} {\bf Input:} 
	\\
	\hspace*{0.4in} ${\bf y}$   \Comment{the received pilot} \\
	\hspace*{0.4in} $L$  \Comment {number of paths}  \\
	\hspace*{0.4in} ${\bf W}$  \Comment {the three dimensional codebook}  \\
	\hspace*{0.02in} {\bf Output:} \\
	\hspace*{0.4in} $ \tilde{\bf h} $  \Comment{the estimated channel} 
	\begin{algorithmic}[1]
		\State Initialization: ${\bf Y} = {\bf y}$, ${\bf \gamma}=\{ \emptyset \}$ 
		\For {$l \in \{ 1,2,\cdots, L \}$}
		\State Calculate the correlation matrix: ${\bf \Gamma} = {\bf W}^{\rm H}{\bf Y}$
		\State Detect new support: $p^{*} = {\rm argmax}_p \left| {\bf \Gamma}_p \right| $ 
		\State Update support set: $\gamma = \gamma \cup p^{*}$
		\State Pseudo inverse: ${\bf W}^{\dag} = ({\bf W}_{:,\gamma}^{\rm T}{\bf W}_{:,\gamma})^{-1}{\bf W}_{:,\gamma}^{\rm T}$
		\State Orthogonal projection: ${\bf h}^{\rm P} = {\bf W}^{\dag}{\bf y}$
		\State Update residual: ${\bf Y} = {\bf Y} - {\bf W}_{:,\gamma}{\bf h}^{\rm P}$
		\EndFor
		
		\State $\tilde{\bf h} ={\bf W}_{:,\gamma} {\bf h}^{\rm P} $
		\State \Return $ \tilde{\bf h}$
	\end{algorithmic}
\end{algorithm}

\subsection{Subspace based channel estimation}

From \cite{demir2022channel} it is known that we can estimate a channel by using the subspace of an omni-directional channel model. When isotropic scattering environment is considered, the correlation function at the receiver is assumed to be 
\begin{equation}
	R({{\bf r}_1},{{\bf r}_2}) = {\rm sinc}\left( \frac{2\| {\bf r}_1-{\bf r}_2 \|}{\lambda} \right).
\end{equation}
Then, the coupling matrix ${\bf R}$ is sampled from the correlation function. A compact eigenvalue decomposition is performed on ${\bf R}$ to obtain ${\bf R} = {\bf U}_1{\bf \Lambda}_1{\bf U}_1^{\rm H}$, where ${\bf \Lambda}_1$ contains the non-zero eigenvalues of ${\bf R}$. The channel estimator is expressed by
\begin{equation}
	\tilde{\bf h} = {\bf U}_1{\bf U}_1^{\rm H}{\bf y}/\sqrt{P}.
\end{equation}
In fact, if we further utilize the information contained in the eigenvalues of ${\bf R}$, the estimation precision can be improved, which corresponds to the channel estimator
\begin{equation}
	\tilde{\bf h} = \sqrt{P}{\bf U}_1(P\boldsymbol{\Lambda}_1+{\bf I})^{-1}{\bf U}_1^{\rm H}{\bf y}.
\end{equation}

\subsection{Proposed channel estimation scheme}
We propose a channel estimation scheme based on the sparsity of the channel model. By reshaping channel vector to ${\bf H}_{i,j} = {\bf h}_{(i-1)*n+j}$, we obtain matrix ${\bf H}$ which has sparsity in the wavenumber domain according to Section \ref{sec_charac}. Therefore, we can detect the peaks in the wavenumber domain to capture the directions of the incident waves. Then we can generate an approximate near-field correlation matrix of the electromagnetic field. The generation procedure of the approximated correlation matrix is shown in Algorithm \ref{alg:NFS}. After obtaining the approximate correlation matrix, we then use the following estimator $\tilde{\bf h} = \sqrt{P}\hat{\bf R}(P \hat{\bf R}+{\bf I})^{-1}{\bf y}$.

\begin{algorithm}[t]
	\caption{NFS correlation function generator} 
	\label{alg:NFS}
	\hspace*{0.02in} {\bf Input:} 
	\\
	\hspace*{0.4in} ${\bf y}$   \Comment{the received pilot} \\
	\hspace*{0.4in} $N_y, N_z$  \Comment {number of antennas}  \\
	\hspace*{0.4in} $d_y,d_z$  \Comment {antenna spacing}  \\
	\hspace*{0.4in} ${\bf F}_1, {\bf F}_2$  \Comment {Fourier transform matrix}  \\
	\hspace*{0.4in} $\lambda$  \Comment {wavelength}  \\
	\hspace*{0.4in} $\eta$   \Comment{threshold}  \\  
	\hspace*{0.4in} $d, r, a, \boldsymbol{\mu}$    \Comment {fixed parameters for simplicity}\\
	\hspace*{0.02in} {\bf Output:} \\
	\hspace*{0.4in} $ \hat{\bf{R}} $  \Comment{the constructed correlation matrix} 
	\begin{algorithmic}[1]
		\State Vector to matrix: ${\bf Y}_{i,:} = {\bf y}_{(i-1)*N_z+1:i*N_z,1}^{\rm T}$ 
		\State Fourier transform on both sides: ${\bf Y}' = {\bf F}_1{\bf Y}{\bf F}_2^{\rm H}$
		\State Average value: $\bar{|{\bf Y}'|} = {\rm sum}\left(|{\bf Y}'|\right)/N_yN_z$
		\State Initialization: ${\bf R} = {\rm zeros}(N_yN_z,N_yN_z), \Phi = \emptyset$
		\State Peak value selection: 
		\For {$i = 1:N_y$}
		\For {$j = 1:N_z$}
		\If{$|{\bf Y}'_{i,j}| > {\rm max}\left(|{\bf Y}'_{i \pm 1,j \pm 1}|, \eta \bar{|{\bf Y}'|}\right)$}
		\State $k_y = \frac{N_y-1-2*i}{N_y-1}$
		\State $k_z = \frac{2*j-N_z+1}{N_z-1}$
		\State ${\bf d} = d*[\sqrt{\frac{2\pi}{\lambda}^2-k_y^2-k_z^2},k_y,k_z]^{\rm T}$
		\State $\Phi = \Phi \cup \{{\bf d}\}$
		\EndIf
		\EndFor
		\EndFor
		
		\For {$i = 1:N_yN_z$}
		\For {$j = 1:N_yN_z$}
		\State ${\bf r}_1 = \left[0, \lfloor \frac{i-1}{N_z} \rfloor-\frac{N_y-1}{2}, (i-1)\%N_z- \frac{N_z-1}{2} \right]$
		\State ${\bf r}_2 = \left[0, \lfloor \frac{j-1}{N_z} \rfloor-\frac{N_y-1}{2}, (j-1)\%N_z- \frac{N_z-1}{2} \right]$
		\State Generate the correlation function $R$ according \\ \ \ \ \ \ \ \ \ \ to {\bf Lemma 1} based on $\Phi$ 
		\State $\hat{\bf R}_{i,j} = R({\bf r}_1,{\bf r}_2)$
		\EndFor
		\EndFor

		\State \Return $ \hat{\bf R}$
	\end{algorithmic}
\end{algorithm}

Here for simplicity we only design the correlation matrix based on the estimated incident wave direction. 
Obviously, the approximation of the correlation matrix can be improved by further estimating or optimizing the parameters $r$, $a$ and $\boldsymbol{\mu}$. In fact, the estimation error when using an approximated correlation matrix ${\bf R}'$ instead of the true correlation matrix ${\bf R}$ can be expressed by the following lemma:

\begin{lemma}[Estimated error when using the proposed scheme]
	The estimated error when using an approximated correlation matrix is $\mathbb{E}\left[ (\tilde{\bf h}-{\bf h})^{\rm H} (\tilde{\bf h}-{\bf h})\right] = {\rm tr}\Big( P (P {\bf I}+\hat{\bf R}^{-1})^{-1}(P {\bf I}+\hat{\bf R}^{-1})^{-1} (P{\bf R}+{\bf I})
	- 2P{\bf R}(P{\bf I}+\hat{\bf R}^{-1})^{-1}+{\bf R}	   \Big)$.
\end{lemma}
\begin{IEEEproof}
	See Appendix C.
\end{IEEEproof}
\begin{corollary}
	When $P \rightarrow 0$, the estimated error will approach ${\rm tr}({\bf R})$. When $P \rightarrow \infty$, the estimated error will approach $0$ whatever the approximate correlation matrix $\hat{\bf R}$ is. Therefore, the performance limit with extremly high or low SNR does not depend on the choice of $\hat{\bf R}$. However, for general $P$, the approximation error will reach the minimum value when $\hat{\bf R} = {\bf R}$, which corresponds to the classical minimum mean square error (MMSE) channel estimator with full information of the distribution of the electromagnetic fields. 
\end{corollary}

\begin{figure}
	\centering 
	\includegraphics[width=0.5\textwidth]{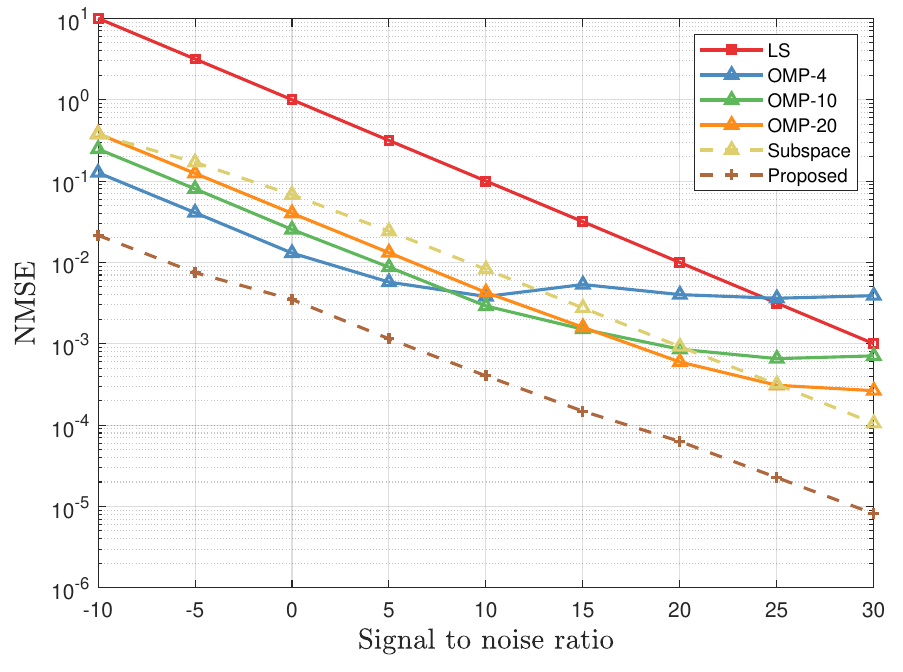} 
	\caption{Comparison of different channel estimation schemes for a $41 \times 41$ antenna array with $
		\lambda/8$ antenna spacing. The proposed channel model is used. } 
	\label{fig_channel_estimation}
\end{figure}

In Fig. \ref{fig_channel_estimation} we have shown the performance comparison between different channel estimation schemes shown above. Here we use the proposed near-field channel model to generate the channel realizations, where four scattering regions exist in the space. We set the antenna array to be $41\times 41$ with $\lambda/8$ antenna spacing. The wavelength of the electromagnetic field is set to $0.2\,{\rm m}$. Specifically, we plot the normalized mean square error (NMSE) $\frac{\| \tilde{\bf h} -{\bf h} \|}{\| {\bf h} \|}$ with the change of SNR $P$. It can be observed that the proposed scheme outperforms traditional schemes like OMP or subspace based channel estimation scheme. For example, when NMSE equals $10^{-3}$, the proposed scheme achieves $12{\rm dB}$ performance gain compared to OMP scheme whose support set is 20. The reason that the proposed scheme can outperform existing schemes can be explained as follows. For the subspace based channel estimation scheme, it considers the incident waves from all directions, which covers the full wavenumber domain. On the contrary, the proposed scheme focuses on a smaller region in the wavenumber domain, thus providing a better approximation of the true correlation function. For the OMP scheme, note that it highly relies on the lattice points of the electromagnetic field, it may not behave well in high SNR region for ${\bf h}$ generated from the correlation function in continuous space.

\begin{figure}
	\centering 
	\includegraphics[width=0.5\textwidth]{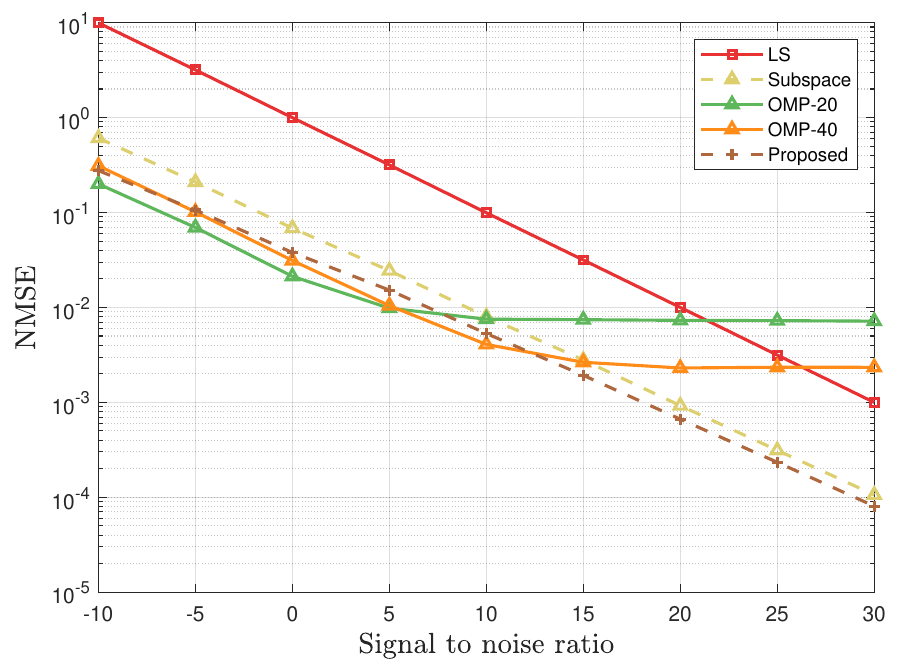} 
	\caption{Comparison of different channel estimation schemes for a $81\times 81$ antenna array with $
		\lambda/8$ antenna spacing. The CDL-D channel model is used.} 
	\label{fig_CDL_channel_estimation}
\end{figure}

Furthermore, we have applied the proposed scheme on CDL-D channel instead of the channel generated by our correlation matrix to further verify its correctness. We adopt a CDL-D channel model which has $81\times 81$ size receiver array with $\lambda/8$ antenna spacing. The wavenumber of the electromagnetic field is set to be $0.4\,{\rm m}$. The simulation result is shown in Fig. \ref{fig_CDL_channel_estimation}. It can be observed that the proposed channel estimation algorithm can also work under classical channel model and outperform existing algorithms like LS and subspace based channel estimation algorithm. For OMP algorithm, its performance is better than the proposed algorithm when SNR is lower than $10\,{\rm dB}$. However, it will still face error platform in the high SNR region, which can be solved by using the proposed algorithm. For example, the proposed scheme can achieve 5dB performance gain compared to 40-points OMP when NMSE is fixed to $2\times10^{-3}$.

\section{Conclusions}
\label{Sec_conclusion}
In this paper, we propose the near-field channel modeling scheme for EIT based on electromagnetic scattering theory. Then, we derive the analytical expression of the correlation function of the fields and analyze the characteristics of it. The proposed scheme can provide a more accurate analytical channel model for EIT than the existing works. Finally, we design a channel estimation scheme for near-field scenario. Numerical analysis verifies the correctness of the proposed scheme and shows that it can outperform existing schemes like LS, OMP, and subspace based channel estimation schemes. Under CDL channel model, the proposed scheme can achieve 5 dB performance gain compared to 40-points OMP when NMSE is fixed to $2\times10^{-3}$.

Further work can be done by integrating the proposed model and traditional near-field model where some scatterers are invisible to part of the array.  

   \section*{Appendix A \\ Proof of Lemma 1}
   Based on the assumption that $r_s = {\rm max} \rho \ll d$, we can use the Taylor expansion to simplify $\|{\bf r}_1-{\bf r}' \|$ and $\|{\bf r}_2-{\bf r}' \|$. First the item $\left( \frac{\rho}{d} \right)^2$ can be ignored. Then we can approximate $\| {\bf r}-{\bf r}' \|$ by $d\left( \sqrt{A({\bf r})} + \frac{\rho}{d}\frac{B({\bf r},\hat{\boldsymbol{\rho}})}{\sqrt{A({\bf r})}} \right)$. The Green's function $g({\bf r},{\bf r}')$ has amplitude item $\frac{1}{4\pi\| {\bf r}-{\bf r}'\|}$ and phase item $e^{{\rm j}k\| {\bf r}-{\bf r}' \|}$. For the distance item we further approximate it by $\frac{1}{4\pi d\sqrt{A({\bf r})}}$. Then, the correlation function of the received field can be approximated by 
   \ifx\onecol\undefined
   \begin{equation}
	   \begin{aligned}
		   &R({\bf r}_1,{\bf r}_2) \approx \tilde{R}({\bf r}_1,{\bf r}_2) =\beta \int_{V} \frac{1}{16\pi^2d^2\sqrt{A({\bf r}_1)A({\bf r}_2)}}\\&~~~~e^{{\rm j}\frac{2\pi}{\lambda}R \left(\sqrt{A({\bf r}_1)}-\sqrt{A({\bf r}_2)}\right)} e^{{\rm j}\frac{2\pi}{\lambda} \rho \left( \frac{B({\bf r}_1,\hat{\boldsymbol{\rho}})}{\sqrt{A({\bf r}_1)}} - \frac{B({\bf r}_2,\hat{\boldsymbol{\rho}})}{\sqrt{A({\bf r}_2)}} \right) }  f({\bf r}'){\rm d}{\bf r}',
	   \end{aligned}
	   \label{equ_Rmn_3d_approx}
   \end{equation}
   \else
   \begin{equation}
	   \begin{aligned}
		R({\bf r}_1,{\bf r}_2) \approx \tilde{R}({\bf r}_1,{\bf r}_2) =\beta \int_{V} \frac{1}{16\pi^2d^2\sqrt{A({\bf r}_1)A({\bf r}_2)}}e^{{\rm j}\frac{2\pi}{\lambda}R \left(\sqrt{A({\bf r}_1)}-\sqrt{A({\bf r}_2)}\right)} e^{{\rm j}\frac{2\pi}{\lambda} \rho \left( \frac{B({\bf r}_1,\hat{\boldsymbol{\rho}})}{\sqrt{A({\bf r}_1)}} - \frac{B({\bf r}_2,\hat{\boldsymbol{\rho}})}{\sqrt{A({\bf r}_2)}} \right) }  f({\bf r}'){\rm d}{\bf r}',
	   \end{aligned}
	   \label{equ_Rmn_3d_approx}
   \end{equation}
   \fi
   where $f({\bf r}'){\rm d}{\bf r}' = \frac{a+1}{\pi r_s^{2a+2}} \rho (r_s^2-\rho^2)^{a} {\rm d}\rho {\rm d}\theta$. Specifically, if $a =0$, we have uniform distribution on the scatterer, where $f({\bf r}'){\rm d}{\bf r}' = \frac{\rho}{\pi r_s^2} {\rm d}\rho {\rm d}\theta$. 
   
   For the simple case with uniform distribution on the circle, we have 
   \begin{equation}
	   \begin{aligned}
		   \tilde{R}({\bf r}_1,{\bf r}_2) =& \beta \frac{1}{16\pi^2d^2\sqrt{A({\bf r}_1)A({\bf r}_2)}}e^{{\rm j}\frac{2\pi}{\lambda}R \left(\sqrt{A({\bf r}_1)}-\sqrt{A({\bf r}_2)}\right)} \\&\int_0^{r_s} \int_0^{2\pi} e^{{\rm j}\frac{2\pi}{\lambda} \rho \left( \frac{B({\bf r}_1,\hat{\boldsymbol{\rho}})}{\sqrt{A({\bf r}_1)}} - \frac{B({\bf r}_2,\hat{\boldsymbol{\rho}})}{\sqrt{A({\bf d}_2)}} \right) } \frac{\rho}{\pi r_s^2} {\rm d}\theta {\rm d}\rho.
		\end{aligned}
	\end{equation}
	We first focus on the integral on the angle $\theta$. Since $B({\bf r},\hat{\boldsymbol{\rho}})$ contains the exponentionals of $\cos\theta$ and $\sin \theta$, we adopt the \cite[Eq. (3.937)]{table} which shows that 
	\begin{equation}
		\begin{aligned}
			&~~~~\int_0^{2\pi} e^{p\cos x+q\sin x}e^{{\rm j}(a\cos x+b\sin x-mx)}{\rm d}x \\&= 2\pi [(b-p)^2+(a+q)^2]^{-\frac{m}{2}}(A-{\rm j}B)^{\frac{m}{2}}I_m(\sqrt{C+{\rm j}D}),
		\end{aligned}
	\end{equation}	
	where $A = p^2-q^2+a^2-b^2$, $B=2pq+2ab$, $C=p^2+q^2-a^2-b^2$ and $D = -2ap-2bq$. Comparing with the integral, we have $p = q = m = 0$, $a = \frac{2\pi}{\lambda}\rho \Bigg( \frac{\hat{\bf d}\cdot \hat{\boldsymbol{\mu}}_1}{\sqrt{A({\bf r}_1)}} - \frac{\hat{\bf d}\cdot \hat{\boldsymbol{\mu}}_1}{\sqrt{A({\bf r}_2)}} - \frac{r_1}{d} \frac{\hat{\bf r}_1 \cdot \hat{\boldsymbol{\mu}}_1}{\sqrt{A({\bf r}_1)}} + \frac{r_2}{R} \frac{\hat{\bf r}_2 \cdot \hat{\boldsymbol{\mu}}_1}{\sqrt{A({\bf r}_2)}}\Bigg)$, $b = -\frac{2\pi}{\lambda} \rho \Bigg( \frac{\hat{\bf d}\cdot \hat{\boldsymbol{\mu}}_2}{\sqrt{A({\bf r}_1)}} - \frac{\hat{\bf d}\cdot \hat{\boldsymbol{\mu}}_2}{\sqrt{A({\bf r}_2)}} - \frac{r_1}{d} \frac{\hat{\bf r}_1 \cdot \hat{\boldsymbol{\mu}}_2}{\sqrt{A({\bf r}_1)}} + \frac{r_2}{d} \frac{\hat{\bf r}_2 \cdot \hat{\boldsymbol{\mu}}_2}{\sqrt{A({\bf r}_2)}}\Bigg)$. Then, we can obtain
	\begin{equation}
		\begin{aligned}
		   &\tilde{R}({\bf r}_1,{\bf r}_2) =  \beta \frac{1}{16\pi^2d^2\sqrt{A({\bf r}_1)A({\bf r}_2)}}e^{{\rm j}\frac{2\pi}{\lambda}R \left(\sqrt{A({\bf r}_1)}-\sqrt{A({\bf r}_2)}\right)} \\&\frac{2}{r_s^2} \int_0^{r_s} I_0({\rm j}\rho \sqrt{C}) \rho {\rm d}\rho,
	   \end{aligned}
   \end{equation}

   where 
   \ifx\onecol\undefined
   \begin{equation}
	   \begin{aligned}
		   C =& \left( \frac{2\pi}{\lambda}  \right)^2 \Bigg( \frac{\hat{\bf d}\cdot \hat{\boldsymbol{\mu}}_1}{\sqrt{A({\bf r}_1)}} - \frac{\hat{\bf d}\cdot \hat{\boldsymbol{\mu}}_1}{\sqrt{A({\bf r}_2)}} - \frac{r_1}{d} \frac{\hat{\bf r}_1 \cdot \hat{\boldsymbol{\mu}}_1}{\sqrt{A({\bf r}_1)}} \\&~~~~~~~~~~~~+ \frac{r_2}{R} \frac{\hat{\bf r}_2 \cdot \hat{\boldsymbol{\mu}}_1}{\sqrt{A({\bf r}_2)}}\Bigg)^2 \\&+  \left( \frac{2\pi}{\lambda}  \right)^2 \Bigg( \frac{\hat{\bf d}\cdot \hat{\boldsymbol{\mu}}_2}{\sqrt{A({\bf r}_1)}} - \frac{\hat{\bf d}\cdot \hat{\boldsymbol{\mu}}_2}{\sqrt{A({\bf r}_2)}} - \frac{r_1}{d} \frac{\hat{\bf r}_1 \cdot \hat{\boldsymbol{\mu}}_2}{\sqrt{A({\bf r}_1)}} \\&~~~~~~~~~~~~+ \frac{r_2}{d} \frac{\hat{\bf r}_2 \cdot \hat{\boldsymbol{\mu}}_2}{\sqrt{A({\bf r}_2)}}\Bigg)^2.
	   \end{aligned}
   \end{equation}
   \else
   \begin{equation}
	   \begin{aligned}
		C =& \left( \frac{2\pi}{\lambda}  \right)^2 \Bigg( \frac{\hat{\bf d}\cdot \hat{\boldsymbol{\mu}}_1}{\sqrt{A({\bf r}_1)}} - \frac{\hat{\bf d}\cdot \hat{\boldsymbol{\mu}}_1}{\sqrt{A({\bf r}_2)}} - \frac{r_1}{d} \frac{\hat{\bf r}_1 \cdot \hat{\boldsymbol{\mu}}_1}{\sqrt{A({\bf r}_1)}} \\&~~~~~~~~~~~~+ \frac{r_2}{R} \frac{\hat{\bf r}_2 \cdot \hat{\boldsymbol{\mu}}_1}{\sqrt{A({\bf r}_2)}}\Bigg)^2 \\&+  \left( \frac{2\pi}{\lambda}  \right)^2 \Bigg( \frac{\hat{\bf d}\cdot \hat{\boldsymbol{\mu}}_2}{\sqrt{A({\bf r}_1)}} - \frac{\hat{\bf d}\cdot \hat{\boldsymbol{\mu}}_2}{\sqrt{A({\bf r}_2)}} - \frac{r_1}{d} \frac{\hat{\bf r}_1 \cdot \hat{\boldsymbol{\mu}}_2}{\sqrt{A({\bf r}_1)}} \\&~~~~~~~~~~~~+ \frac{r_2}{d} \frac{\hat{\bf r}_2 \cdot \hat{\boldsymbol{\mu}}_2}{\sqrt{A({\bf r}_2)}}\Bigg)^2.
	   \end{aligned}
   \end{equation}
   \fi
   
   Then, according to \cite[Eq. (6.561)]{table} we have 
   \begin{equation}
	   \begin{aligned}
		   \tilde{R}({\bf r}_1,{\bf r}_2) =& \frac{\beta}{16\pi^2d^2\sqrt{A({\bf r}_1)A({\bf r}_2)}}e^{{\rm j}\frac{2\pi}{\lambda}R \left(\sqrt{A({\bf r}_1)}-\sqrt{A({\bf r}_2)}\right)} \\&\frac{2}{r_s^2} \int_0^{r_s} J_0(\rho \sqrt{C}) \rho {\rm d}\rho
		   \\ =& \frac{\beta}{8\pi^2d^2\sqrt{A({\bf r}_1)A({\bf r}_2)}}e^{{\rm j}\frac{2\pi}{\lambda}R \left(\sqrt{A({\bf r}_1)}-\sqrt{A({\bf r}_2)}\right)}  \\&\frac{1}{r_s\sqrt{C}}J_1(\sqrt{C}r_s).
	   \end{aligned}
   \end{equation}		
   
   If we adopt $f({\bf r}'){\rm d}{\bf r}' = \frac{a+1}{\pi r_s^{2a+2}} \rho (r_s^2-\rho^2)^{a} {\rm d}\rho {\rm d}\theta$, according to \cite[Eq. (6.567)]{table} we have 
   \begin{equation}
	   \begin{aligned}
		   \tilde{R}({\bf r}_1,{\bf r}_2) =& \frac{\beta}{16\pi^2d^2\sqrt{A({\bf r}_1)A({\bf r}_2)}}e^{{\rm j}\frac{2\pi}{\lambda}R \left(\sqrt{A({\bf r}_1)}-\sqrt{A({\bf r}_2)}\right)} \\&2\pi \frac{a+1}{\pi r_s^{2a+2}} \int_0^{r_s} J_0(\rho \sqrt{C}) \rho (r_s^2-\rho^2)^{a} {\rm d}\rho
		   \\ =& \frac{\beta}{16\pi^2d^2\sqrt{A({\bf r}_1)A({\bf r}_2)}}e^{{\rm j}\frac{2\pi}{\lambda}R \left(\sqrt{A({\bf r}_1)}-\sqrt{A({\bf r}_2)}\right)} \\&\frac{2(a+1)}{r_s^{2a+2}} r_s^{2a+2} \int_0^1 J_0(\rho' r_s \sqrt{C})\rho' (1-\rho')^2 {\rm d}\rho'
		   \\ =& \frac{\beta}{16\pi^2d^2\sqrt{A({\bf r}_1)A({\bf r}_2)}}e^{{\rm j}\frac{2\pi}{\lambda}R \left(\sqrt{A({\bf r}_1)}-\sqrt{A({\bf r}_2)}\right)} \\&2(a+1) 2^{a} \Gamma (a+1) (\sqrt{C}r_s)^{-(a+1)} J_{a+1} (\sqrt{C}r_s).
	   \end{aligned}
	   \label{equ_correlation_approximated}
   \end{equation}

   \section*{Appendix B \\ Proof of Lemma 2}
   We have 
   \begin{equation}
	   \begin{aligned}
   &~~~~\|{\bf d}+\boldsymbol{\rho}-{\bf r}\| \\&= \sqrt{d^2+\rho^2+r^2+2{\bf d}\cdot\boldsymbol{\rho}-2{\bf d}\cdot{\bf r}-2\boldsymbol{\rho}\cdot{\bf r}}
   \\& = d \sqrt{1+\left( \frac{\rho}{d} \right)^2 + \left( \frac{r}{d} \right)^2+2\frac{{\bf d}\cdot{\bf \rho}}{d^2}-2\frac{{\bf d}\cdot{\bf r}}{d^2}-2\frac{\boldsymbol{\rho}\cdot{\bf r}}{d^2}}.
   \end{aligned}
   \end{equation}
   According to the Taylor expansion $\sqrt{1+x} \approx 1+\frac{1}{2}x-\frac{1}{8}x^2 $ and the assumption that $d \gg {\rm max}(r,\rho)$, we have that 
   \begin{equation}
	   \begin{aligned}
   &~~~~\|{\bf d}+\boldsymbol{\rho}-{\bf r}\| \\&\approx d\Bigg( 1+ \frac{{\bf d}\cdot\boldsymbol{\rho}}{d^2}-\frac{{\bf d}\cdot{\bf r}}{d^2}+ \frac{\rho^2}{2d^2}  +\frac{r^2}{2d^2} -\frac{\boldsymbol{\rho}\cdot{\bf r}}{d^2} \\&~~~~-\frac{({\bf d}\cdot\boldsymbol{\rho})^2}{2d^4} - \frac{({\bf d}\cdot{\bf r})^2}{2d^4} + \frac{({\bf d}\cdot\boldsymbol{\rho})({\bf d}\cdot{\bf r})}{d^4} \Bigg),
   \end{aligned}
   \end{equation}
   where higher orders of $\frac{\rho}{d}$ and $\frac{r}{d}$ are neglected. For far-field channel modeling without considering the size of scatterers, only the terms $d\left( 1- \frac{{\bf d}\cdot{\bf r}}{d^2}  \right)$ is kept, which means that the rest terms should be small enough. If we adopt the $\frac{\pi}{8}$ phase error as the threshold, we have 
   {
   \begin{equation}
	   \begin{aligned}
   &\frac{2\pi}{\lambda} \Bigg| \frac{{\bf d}\cdot\boldsymbol{\rho}}{d} +\frac{\rho^2}{2d} + \frac{r^2}{2d} - \frac{\boldsymbol{\rho}\cdot{\bf r}}{d}- \frac{({\bf d}\cdot\boldsymbol{\rho})^2}{2d^3} - \frac{({\bf d}\cdot{\bf r})^2}{2d^3} \\&+ \frac{({\bf d}\cdot\boldsymbol{\rho})({\bf d}\cdot{\bf r})}{d^3} \Bigg|
   \\& \leqslant \frac{2\pi}{\lambda} \left( r_{\rm s}+\frac{(r_{\rm s}+r_{\rm m})^2}{2d} \right) \leqslant \frac{\pi}{8},
	   \end{aligned}
   \end{equation}
   which leads to $(\lambda-16r_{\rm s})d \geqslant 8(r_{\rm s}+r_{\rm m})^2 $. Therefore, when the radius $r_{\rm s}$ of the scatterer is larger than $\frac{\lambda}{16}$ and $d\geqslant \frac{8(r_{\rm s}+r_{\rm m})^2}{\lambda-16r_s}$, the scatterer size has to be taken into consideration.} Moreover, if we keep the terms $d \left( 1+\frac{{\bf d}\cdot\boldsymbol{\rho}}{d^2} - \frac{{\bf d}\cdot{\bf r}}{d^2} \right)$, which means that the antenna array and the scatterer are in each other's far-field respectively, we have
   {
   \begin{equation}
	   \begin{aligned}
   &\frac{2\pi}{\lambda} \Bigg| \frac{\rho^2+r^2}{2d} - \frac{\boldsymbol{\rho}\cdot{\bf r}}{d}- \frac{({\bf d}\cdot\boldsymbol{\rho})^2}{2d^3} - \frac{({\bf d}\cdot{\bf r})^2}{2d^3} + \frac{({\bf d}\cdot\boldsymbol{\rho})({\bf d}\cdot{\bf r})}{2d^3} \Bigg|
   \\& \leqslant \frac{2\pi}{\lambda} \left( \frac{(r_{\rm s}+r_{\rm m})^2}{d} \right) \leqslant \frac{\pi}{8},
	   \end{aligned}
   \end{equation}
   which leads to $d\geqslant \frac{8(r_{\rm s}+r_{\rm m})^2}{\lambda}$. 
}
   
   \section*{Appendix C \\ Proof of Lemma 3}
   Note that the channel ${\bf h}$ is a random vector whose distribution is controlled by its correlation matrix ${\bf R}$.
   For the difference between the estimated channel $\tilde{\bf h}$ and the true channel ${\bf h}$, we have 
   \begin{equation}
   	\begin{aligned}
   		&\mathbb{E}\left[ (\tilde{\bf h}-{\bf h})^{\rm H} (\tilde{\bf h}-{\bf h})\right] \\=& \mathbb{E} \left[ (\sqrt{P}\hat{\bf R}(P \hat{\bf R}+{\bf I})^{-1}{\bf y}-{\bf h})^{\rm H} (\sqrt{P}\hat{\bf R}(P \hat{\bf R}+{\bf I})^{-1}{\bf y} -{\bf h}) \right]
   		\\=& \mathbb{E}\left[ {\rm tr}(\sqrt{P}\hat{\bf R}(P \hat{\bf R}+{\bf I})^{-1}{\bf y}-{\bf h}) (\sqrt{P}\hat{\bf R}(P \hat{\bf R}+{\bf I})^{-1}{\bf y} -{\bf h})^{\rm H} \right] 
   		\\=&  {\rm tr}\Big(\sqrt{P}\hat{\bf R}(P \hat{\bf R}+{\bf I})^{-1})(P{\bf R}+{\bf I})(\sqrt{P}\hat{\bf R}(P \hat{\bf R}+{\bf I})^{-1})^{\rm H} 
   		\\&- \sqrt{P}\hat{\bf R}(P \hat{\bf R}+{\bf I})^{-1}\sqrt{P}{\bf R}\\&-\sqrt{P}{\bf R}(\sqrt{P}\hat{\bf R}(P \hat{\bf R}+{\bf I})^{-1})^{\rm H}
   		 + {\bf R} \Big)
   		\\ = & {\rm tr}\Big( P (P \hat{\bf R}+{\bf I})^{-1}{\hat {\bf R}}{\hat {\bf R}}(P \hat{\bf R}+{\bf I})^{-1} (P{\bf R}+{\bf I})
   		\\& - 2P{\bf R}(P{\bf I}+\hat{\bf R}^{-1})^{-1}+{\bf R}	   \Big)
   		\\ = & {\rm tr}\Big( P (P {\bf I}+\hat{\bf R}^{-1})^{-1}(P {\bf I}+\hat{\bf R}^{-1})^{-1} (P{\bf R}+{\bf I})
   		\\& - 2P{\bf R}(P{\bf I}+\hat{\bf R}^{-1})^{-1}+{\bf R}	   \Big).
   	\end{aligned}
   \end{equation}

\ifCLASSOPTIONcaptionsoff
 \newpage
\fi

\footnotesize

\bibliographystyle{IEEEtran}

\bibliography{near_field_EIT}

\begin{thebibliography}{10}
\providecommand{\url}[1]{#1}
\csname url@samestyle\endcsname
\providecommand{\newblock}{\relax}
\providecommand{\bibinfo}[2]{#2}
\providecommand{\BIBentrySTDinterwordspacing}{\spaceskip=0pt\relax}
\providecommand{\BIBentryALTinterwordstretchfactor}{4}
\providecommand{\BIBentryALTinterwordspacing}{\spaceskip=\fontdimen2\font plus
\BIBentryALTinterwordstretchfactor\fontdimen3\font minus
  \fontdimen4\font\relax}
\providecommand{\BIBforeignlanguage}[2]{{%
\expandafter\ifx\csname l@#1\endcsname\relax
\typeout{** WARNING: IEEEtran.bst: No hyphenation pattern has been}%
\typeout{** loaded for the language `#1'. Using the pattern for}%
\typeout{** the default language instead.}%
\else
\language=\csname l@#1\endcsname
\fi
#2}}
\providecommand{\BIBdecl}{\relax}
\BIBdecl

\bibitem{basar2019wireless}
E.~Basar, M.~Di~Renzo, J.~De~Rosny, M.~Debbah, M.-S. Alouini, and R.~Zhang,
  ``Wireless communications through reconfigurable intelligent surfaces,''
  \emph{IEEE Access}, vol.~7, pp. 116\,753--116\,773, Aug. 2019.

\bibitem{wang2022location}
Z.~Wang, Z.~Liu, Y.~Shen, A.~Conti, and M.~Z. Win, ``Location awareness in
  beyond {5G} networks via reconfigurable intelligent surfaces,'' \emph{{IEEE}
  J. Sel. Areas Commun.}, vol.~40, no.~7, pp. 2011--2025, Jul. 2022.

\bibitem{huang2020holographic}
C.~Huang, S.~Hu, G.~C. Alexandropoulos, A.~Zappone, C.~Yuen, R.~Zhang,
  M.~Di~Renzo, and M.~Debbah, ``Holographic {MIMO} surfaces for {6G} wireless
  networks: Opportunities, challenges, and trends,'' \emph{IEEE Wireless
  Commun.}, vol.~27, no.~5, pp. 118--125, Oct. 2020.

\bibitem{zhang2023pattern}
Z.~Zhang and L.~Dai, ``Pattern-division multiplexing for multi-user
  continuous-aperture {MIMO},'' \emph{IEEE J. Sel. Areas Commun.}, vol.~41,
  no.~8, pp. 2350--2366, Aug. 2023.

\bibitem{cui2022near}
M.~Cui, Z.~Wu, Y.~Lu, X.~Wei, and L.~Dai, ``Near-field {MIMO} communications
  for {6G}: Fundamentals, challenges, potentials, and future directions,''
  \emph{IEEE Comm. Mag.}, vol.~61, no.~1, pp. 40--46, Jan. 2023.

\bibitem{wu2023multiple}
Z.~Wu and L.~Dai, ``Multiple access for near-field communications: {SDMA} or
  {LDMA}?'' \emph{{IEEE} J. Sel. Areas Commun.}, vol.~41, no.~6, pp.
  1918--1935, Jun. 2023.

\bibitem{chafii2023twelve}
M.~Chafii, L.~Bariah, S.~Muhaidat, and M.~Debbah, ``Twelve scientific
  challenges for {6G}: Rethinking the foundations of communications theory,''
  \emph{IEEE Comm. Surveys Tut.}, Feb. 2023.

\bibitem{migliore2018horse}
M.~D. Migliore, ``Horse (electromagnetics) is more important than horseman
  (information) for wireless transmission,'' \emph{{IEEE} Trans. Antennas
  Propag.}, vol.~67, no.~4, pp. 2046--2055, Apr. 2018.

\bibitem{zhu2022electromagnetic}
J.~Zhu, Z.~Wan, L.~Dai, M.~Debbah, and H.~V. Poor, ``Electromagnetic
  information theory: Fundamentals, modeling, applications, and open
  problems,'' \emph{{IEEE} Wireless Commun.}, early access, Jan. 2024, doi:
  10.1109/MWC.019.2200602.

\bibitem{gong2023holographic}
T.~Gong, L.~Wei, C.~Huang, Z.~Yang, J.~He, M.~Debbah, and C.~Yuen,
  ``Holographic {MIMO} communications with arbitrary surface placements:
  Near-field {LoS} channel model and capacity limit,'' \emph{IEEE J. Sel. Areas
  Commun.}, early access, Apr. 2023, doi: 10.1109/JSAC.2024.3389126.

\bibitem{wei2023tri}
L.~Wei, C.~Huang, G.~C. Alexandropoulos, Z.~Yang, J.~Yang, E.~Wei, Z.~Zhang,
  M.~Debbah, and C.~Yuen, ``Tri-polarized holographic {MIMO} surfaces for
  near-field communications: Channel modeling and precoding design,''
  \emph{IEEE Trans. Wireless Commun.}, vol.~22, no.~12, pp. 8828--8842, Dec.
  2023.

\bibitem{pizzo2023wide}
A.~Pizzo, A.~Lozano, S.~Rangan, and T.~L. Marzetta, ``Wide-aperture {MIMO} via
  reflection off a smooth surface,'' \emph{{IEEE} Trans. Wireless Commun.},
  vol.~22, no.~8, pp. 5229--5239, Aug. 2023.

\bibitem{bucci1987spatial}
O.~Bucci and G.~Franceschetti, ``On the spatial bandwidth of scattered
  fields,'' \emph{{IEEE} Trans. Antennas Propag.}, vol.~35, no.~12, pp.
  1445--1455, Dec. 1987.

\bibitem{bucci1989degrees}
O.~M. Bucci and G.~Franceschetti, ``On the degrees of freedom of scattered
  fields,'' \emph{{IEEE} Trans. Antennas Propag.}, vol.~37, no.~7, pp.
  918--926, Jul. 1989.

\bibitem{franceschetti2015landau}
M.~Franceschetti, ``On {Landau’s} eigenvalue theorem and information
  cut-sets,'' \emph{{IEEE} Trans. Inf. Theory}, vol.~61, no.~9, pp. 5042--5051,
  Sep. 2015.

\bibitem{jensen2008capacity}
M.~A. Jensen and J.~W. Wallace, ``Capacity of the continuous-space
  electromagnetic channel,'' \emph{{IEEE} Trans. Antennas Propag.}, vol.~56,
  no.~2, pp. 524--531, Feb. 2008.

\bibitem{jeon2017capacity}
W.~Jeon and S.-Y. Chung, ``Capacity of continuous-space electromagnetic
  channels with lossy transceivers,'' \emph{{IEEE} Trans. Inf. Theory},
  vol.~64, no.~3, pp. 1977--1991, Mar. 2018.

\bibitem{wan2022mutual}
Z.~Wan, J.~Zhu, Z.~Zhang, L.~Dai, and C.-B. Chae, ``Mutual information for
  electromagnetic information theory based on random fields,'' \emph{{IEEE}
  Trans. Commun.}, vol.~71, no.~4, pp. 1982--1996, Feb. 2023.

\bibitem{chew1999waves}
W.~C. Chew, \emph{Waves and fields in inhomogenous media}.\hskip 1em plus 0.5em
  minus 0.4em\relax John Wiley \& Sons, 1999, vol.~16.

\bibitem{pizzo2022fourier}
A.~Pizzo, L.~Sanguinetti, and T.~L. Marzetta, ``Fourier plane-wave series
  expansion for holographic {MIMO} communications,'' \emph{{IEEE} Trans.
  Wireless Commun.}, vol.~21, no.~9, pp. 6890--6905, Sep. 2022.

\bibitem{bjornson2020rayleigh}
E.~Bj{\"o}rnson and L.~Sanguinetti, ``Rayleigh fading modeling and channel
  hardening for reconfigurable intelligent surfaces,'' \emph{{IEEE} Wireless
  Commun. Lett.}, vol.~10, no.~4, pp. 830--834, Apr. 2021.

\bibitem{Marzetta'20}
A.~Pizzo, T.~L. Marzetta, and L.~Sanguinetti, ``Spatially-stationary model for
  holographic {MIMO} small-scale fading,'' \emph{{IEEE} J. Sel. Areas Commun.},
  vol.~38, no.~9, pp. 1964--1979, Sep. 2020.

\bibitem{pizzo2022spatial}
A.~Pizzo, L.~Sanguinetti, and T.~L. Marzetta, ``Spatial characterization of
  electromagnetic random channels,'' \emph{IEEE Open Journal of the
  Communications Society}, vol.~3, pp. 847--866, Apr. 2022.

\bibitem{demir2022channel}
{\"O}.~T. Demir, E.~Bj{\"o}rnson, and L.~Sanguinetti, ``Channel modeling and
  channel estimation for holographic massive {MIMO} with planar arrays,''
  \emph{IEEE Wireless Commun. Lett.}, vol.~11, no.~5, pp. 997--1001, May 2022.

\bibitem{liu2023near}
Y.~Liu, Z.~Wang, J.~Xu, C.~Ouyang, X.~Mu, and R.~Schober, ``Near-field
  communications: A tutorial review,'' \emph{IEEE Open Journal of the
  Communications Society}, pp. 1999--2049, Aug. 2023.

\bibitem{gruber2008new}
F.~K. Gruber and E.~A. Marengo, ``New aspects of electromagnetic information
  theory for wireless and antenna systems,'' \emph{{IEEE} Trans. Antennas
  Propag.}, vol.~56, no.~11, pp. 3470--3484, Nov. 2008.

\bibitem{poon2005degrees}
A.~S. Poon, R.~W. Brodersen, and D.~N. Tse, ``Degrees of freedom in
  multiple-antenna channels: A signal space approach,'' \emph{{IEEE} Trans.
  Inf. Theory}, vol.~51, no.~2, pp. 523--536, Feb. 2005.

\bibitem{danufane2021path}
F.~H. Danufane, M.~Di~Renzo, J.~De~Rosny, and S.~Tretyakov, ``On the path-loss
  of reconfigurable intelligent surfaces: An approach based on {Green}’s
  theorem applied to vector fields,'' \emph{{IEEE} Trans. Commun.}, vol.~69,
  no.~8, pp. 5573--5592, Aug. 2021.

\bibitem{li2018deepnis}
L.~Li, L.~G. Wang, F.~L. Teixeira, C.~Liu, A.~Nehorai, and T.~J. Cui,
  ``{DeepNIS}: Deep neural network for nonlinear electromagnetic inverse
  scattering,'' \emph{{IEEE} Trans. Antennas Propag.}, vol.~67, no.~3, pp.
  1819--1825, Mar. 2019.

\bibitem{franceschetti2006scattering}
G.~Franceschetti and D.~Riccio, \emph{Scattering, natural surfaces, and
  fractals}.\hskip 1em plus 0.5em minus 0.4em\relax Elsevier, 2006.

\bibitem{cabayan1973scattering}
H.~Cabayan and R.~Murphy, ``Scattering of electromagnetic waves by rough
  perfectly conducting circular cylinders,'' \emph{{IEEE} Trans. Antennas
  Propag.}, vol.~21, no.~6, pp. 893--895, Nov. 1973.

\bibitem{osgood1999x}
R.~Osgood~Iii, S.~Sinha, J.~Freeland, Y.~Idzerda, and S.~Bader, ``X-ray
  scattering from magnetically and structurally rough surfaces,'' \emph{Journal
  of magnetism and magnetic materials}, vol. 198, pp. 698--702, Jun. 1999.

\bibitem{dainty1977statistics}
J.~C. Dainty, ``{The} statistics of speckle patterns,'' in \emph{Progress in
  optics}.\hskip 1em plus 0.5em minus 0.4em\relax Elsevier, 1977, vol.~14, pp.
  1--46.

\bibitem{tang1996regions}
K.~Tang, R.~A. Dimenna, and R.~O. Buckius, ``Regions of validity of the
  geometric optics approximation for angular scattering from very rough
  surfaces,'' \emph{International Journal of Heat and Mass Transfer}, vol.~40,
  no.~1, pp. 49--59, Oct. 1996.

\bibitem{ticconi2011models}
F.~Ticconi, L.~Pulvirenti, N.~Pierdicca, and V.~Zhurbenko, ``Models for
  scattering from rough surfaces,'' \emph{Electromagnetic waves}, vol.~10, pp.
  203--226, 2011.

\bibitem{christou2016far}
M.~A. Christou and A.~C. Polycarpou, ``Far-field scattering from an
  electrically small circular aperture in a conducting screen,'' \emph{IEEE
  Trans. EMC}, vol.~59, no.~2, pp. 404--410, Apr. 2016.

\bibitem{mittal2009angle}
A.~Mittal, R.~Bhattacharjee, and B.~Paul, ``Angle and time of arrival
  statistics for a far circular scattering model,'' in \emph{Proc. of
  NCC}.\hskip 1em plus 0.5em minus 0.4em\relax Citeseer, 2009, pp. 141--145.

\bibitem{lu2023near}
Y.~Lu and L.~Dai, ``Near-field channel estimation in mixed {LoS/NLoS}
  environments for extremely large-scale {MIMO} systems,'' \emph{IEEE Trans.
  Commun.}, vol.~71, no.~6, pp. 3694--3707, Jun. 2023.

\bibitem{mercer1909xvi}
J.~Mercer, ``Functions of positive and negative type, and their connection with
  the theory of integral equations,'' \emph{Philos. Trans. Roy. Soc. London},
  vol. 209, no. 441-458, pp. 415--446, Jan. 1909.

\bibitem{CDL}
{3GPP TR}, ``Study on channel model for frequencies from 0.5 to 100 {GHz},''
  \emph{3GPP TR 38.901 version 14.0.0 Release}, Dec. 2019.

\bibitem{schaubach1992ray}
K.~R. Schaubach, N.~J. Davis, and T.~S. Rappaport, ``A ray tracing method for
  predicting path loss and delay spread in microcellular environments,'' in
  \emph{[1992 Proceedings] Vehicular Technology Society 42nd VTS
  Conference-Frontiers of Technology}.\hskip 1em plus 0.5em minus 0.4em\relax
  IEEE, 1992, pp. 932--935.

\bibitem{selvan2017fraunhofer}
K.~T. Selvan and R.~Janaswamy, ``Fraunhofer and {Fresnel} distances: Unified
  derivation for aperture antennas.'' \emph{{IEEE} Antennas and Propag. Mag.},
  vol.~59, no.~4, pp. 12--15, Aug. 2017.

\bibitem{wavethoeyofinformation}
F.~Massimo, \emph{Wave Theory of Information}.\hskip 1em plus 0.5em minus
  0.4em\relax Cambridge, U.K.: Cambridge Univ. Press, 2017.

\bibitem{dardari2020communicating}
D.~Dardari, ``Communicating with large intelligent surfaces: Fundamental limits
  and models,'' \emph{{IEEE} J. Sel. Areas Commun.}, vol.~38, no.~11, pp.
  2526--2537, Nov. 2020.

\bibitem{wang2012generalized}
J.~Wang, S.~Kwon, and B.~Shim, ``Generalized orthogonal matching pursuit,''
  \emph{IEEE Trans. Signal Process.}, vol.~60, no.~12, pp. 6202--6216, Dec.
  2012.

\bibitem{table}
D.~Zwillinger, V.~Moll, I.~Gradshteyn, and I.~Ryzhik, Eds., \emph{Table of
  Integrals, Series, and Products (Eighth Edition)}.\hskip 1em plus 0.5em minus
  0.4em\relax Boston: Academic Press, 2014.

\end{thebibliography}

\end{document}